\documentclass[final,3p,times,authoryear]{elsarticle}

\usepackage{amssymb}

\usepackage[squaren,Gray,cdot]{SIunits}
\usepackage{color}

\newcommand{\rev}{\textcolor{black}}
\usepackage{soul}
\journal{Journal of the Mechanics and Physics of Solids}

\begin{document}

\begin{frontmatter}

\title{Adhesion in soft contacts is minimum beyond a critical shear displacement}

\author[label1]{C. Oliver}
\author[label1]{D. Dalmas}
\author[label1]{J. Scheibert\corref{cor1}}
\cortext[cor1]{Corresponding author, julien.scheibert@cnrs.fr}
\address[label1]{Univ Lyon, CNRS, Ecole Centrale de Lyon, ENTPE, LTDS, UMR5513, 69134 Ecully, France}

\begin{abstract}
The most direct measurement of adhesion is the pull-off force, i.e. the tensile force necessary to separate two solids in contact. For a given interface, it depends on various experimental parameters, including separation speed, contact age and maximum loading force. Here, using smooth contacts between elastomer spheres and rigid plates, we show that the pull-off force also varies if the contact is sheared prior to separation. For shear displacements below a critical value about 10\% of that necessary to yield gross sliding, the pull-off force steadily decreases as shear increases. For larger shear, the pull-off force remains constant, at a residual value 10-15\% of its initial value. Combining force measurements and in situ imaging, we show how the unloading path leading to contact separation is modified by the initial shear. In particular, we find that the residual pull-off force prevails if the contact reaches full sliding during unloading. \rev{Based on those observations, a first modelling attempt of the critical shear displacement is proposed, involving a competition between jump instability and transition to sliding. Overall,} those results offer new insights into the interplay between adhesion and friction, provide new constraints on adhesion measurements and challenge existing adhesive models. They will be useful wherever soft contacts undergo both normal and shear stresses, including tire grip, soft robotics, haptics and animal locomotion.
\end{abstract}

\begin{graphicalabstract}
\includegraphics[width=\textwidth]{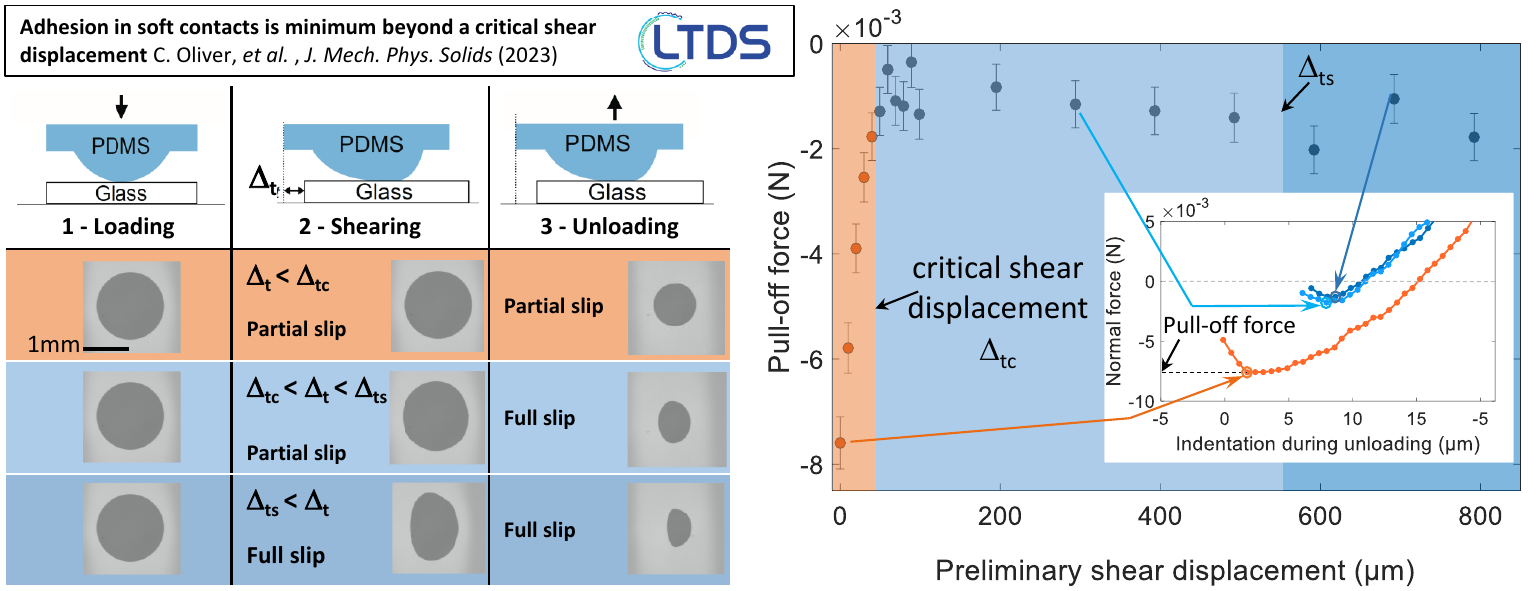}
\end{graphicalabstract}

\begin{highlights}
\item The pull-off force of elastomer contacts decreases with increasing preliminary shear
\item The decrease stops at a critical shear displacement, too small to yield full sliding
\item Beyond, the pull-off force remains constant at about 10\% of its value without shear
\item The constant adhesion regime prevails when full sliding occurs before pull-off
\item \rev{A model for the critical shear displacement is proposed and tested quantitatively}
\end{highlights}

\begin{keyword}
Elastomer \sep Pull-off force \sep Friction \sep Tribology \sep PDMS
\end{keyword}

\end{frontmatter}


\section{Introduction}\label{sec:intro}

Adhesion is the phenomenon by which two solid surfaces tend to stick together, due to attractive inter-atomic or inter-molecular forces~\citep{maugis_contact_2000}. At the continuum level, it is often quantified through the work of adhesion, $w_0$, which is the reversible work necessary to bring a unit area of the interface from contact at equilibrium distance to infinite separation. Adhesion is prominent in contacts involving soft materials, e.g. elastomers, gels and biomaterials, due to their combination of large surface energy and compliance. In related systems, including fingertips~\citep{spinner_sticky_2016}, pressure-sensitive adhesives~\citep{lakrout_direct_1999} and microcontact printing~\citep{carlson_transfer_2012}, adhesion may manifest at macroscale, when measurable tensile normal forces have to be overcome before separating the two solids. The maximum such tensile force is the so-called pull-off force. For smooth contacts between spheres, as described in the classical JKR or DMT models, the pull-off force is proportional to both $w_0$ and the sphere radius~\citep{maugis_contact_2000, barthel_adhesive_2008}. For rough contacts, the pull-off force is often much reduced compared to the smooth case, by an amount depending on the amplitude and spectral contents of the topography~\citep{fuller_effect_1975, pickering_effects_2001, pastewka_contact_2014, vakis_modeling_2018, dalvi_linking_2019}.

Beyond those intrinsic features of the contact interface, the pull-off force is also known to depend on the way the pull-off test is performed. For instance, due to the viscoelastic nature of most soft materials, the pull-off force varies with the separation speed~\citep{violano_jkr-like_2021}. For rough contacts, it also depends on the maximum normal load prior to unloading~\citep{dorogin_role_2017}. In addition, it increases with the time spent in contact before separation~\citep{pickering_effects_2001}. In contrast, one implicit feature of the protocols used for pull-off tests has remained essentially unquestioned: the fact that separation is performed on contacts that have been prepared under a pure normal load. Hence, here, we address the question of what happens in a pull-off test when the contact is shear-loaded prior to separation.

Recent experiments have provided related preliminary answers. First, after full sliding, the pull-off force of an elastomer contact can be significantly reduced compared to that of an un-sheared contact~\citep{peng_effect_2021}. Second, the real contact area under constant normal load decreases as soon as shear is applied to an elastomer interface~\citep{waters_mode-mixity-dependent_2010, sahli_evolution_2018, mergel_continuum_2019}, suggesting that the adhesive behavior of a soft contact is affected by the slightest shear. In this context, we hypothesize that minute shear, well below that necessary to trigger full sliding, may already significantly reduce the pull-off force of a soft contact. In order to test this assumption, in the following, we perform pull-off tests on elastomer contacts submitted to increasing preliminary shear displacements, while all other system and protocol features are kept unchanged.

\section{Methods}\label{sec:methods}
\subsection{Principles of the experiments}
\begin{figure}[hbt!]
\centering
\includegraphics[width=0.7\linewidth]{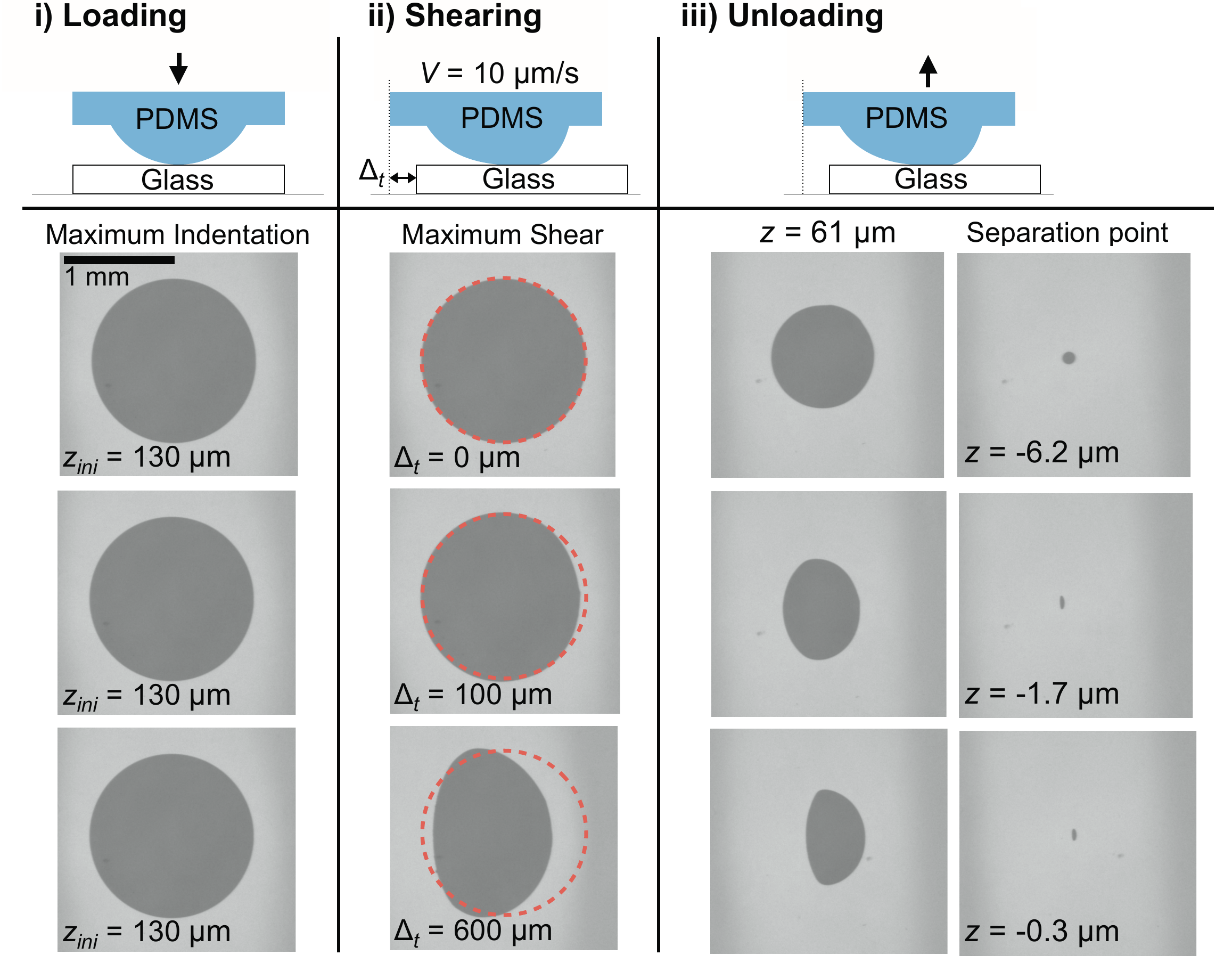}
\caption{Contact loading history and real contact images. Top line: sketch of the three phases of contact: loading, shearing and unloading (columns 1 to 3, respectively). Bottom line: typical images of a standard-PDMS/glass contact, for each phase and for three different preliminary shear displacements, $\Delta_t$ (value indicated). $P_{ini}\simeq$0.1\,N. All images share the same pixel size. Column 1: at the end of the loading phase. Column 2: at the end of the shearing phase. Dashed red circle: contour of the corresponding contact in column 1. Column 3: at two instants during unloading. Left sub-column: at normal indentation \unit{61}{\micro\meter}. Right sub-column: last image before contact separation (corresponding normal indentation indicated).}\label{fig1}
\end{figure}

As sketched in Fig.~\ref{fig1}, we consider smooth dry contacts between a centimetric sphere made of PolyDiMethylSiloxane (PDMS) elastomer and a  glass plate. The contacts are submitted to three successive phases of motion: first, an indentation along the direction normal to the plate (loading phase), reaching a normal load $P_{ini}$; second, a continuous tangential displacement at a velocity $V$=0.01\,mm/s, and of amplitude $\Delta_t$ between 0 and about 1100$\mu$m (shearing phase) ; third, a step-based, displacement-controlled separation of the solids (unloading phase). During all phases, we monitor the evolutions of the normal and tangential forces, $P$ and $Q$, and of the real contact, via in situ imaging.

\subsection{Sample preparation}\label{sec:sample}
All PDMS samples are made of Sylgard 184. They are molded in a concave optical glass lens to produce smooth spherical elastomer caps (curvature radius $R$=\unit{9.42}{\milli\meter}) on top of a bulk (maximum thickness \unit{7}{\milli\meter}, lateral size \unit{21}{\milli\meter}) fixed on a rigid sample holder. For "standard-PDMS" (resp. "soft-PDMS") samples, the mixing mass ratio of base to curing agent is 10:1 (resp. 20:1). For both, the cross-linking protocol is that recommended and justified in~\cite{delplanque_solving_2022}: \unit{24}{\hour} at \unit{25}{\degreecelsius}, demolding, and  \unit{24}{\hour} at \unit{50}{\degreecelsius}. The counter-surface is a smooth glass or PMMA plate, cleaned by wiping with ethanol and drying in ambient air during \unit{5}{\minute}.

\subsection{Mechanical test}
The experiments are performed in a clean room (temperature 20$\pm$1\unit{\degreecelsius}, humidity 50$\pm$10\%), using the setup described in Fig.~13 of~\cite{guibert_versatile_2021}. This opto-mechanical device enables studying the response of a soft contact interface submitted to combined normal and tangential stimuli, by simultaneously measuring the six force/torque components applied to the contact and recording in situ images of the interface. The protocol follows the three steps sketched in Fig.~\ref{fig1}. First, starting from the origin of the vertical displacement ($z$=0, defined as the altitude at which the PDMS sphere first snaps into contact with the plate), the loading phase consists of a rapid normal displacement up to the indentation necessary to reach the target initial normal load $P_{ini}$ ($\simeq$ 0.1, 0.2 or \unit{0.4}{\newton}) without overshooting. The corresponding vertical position is denoted as $z_{ini}$.  The contact is then left to age for \unit{60}{\second}, during which most of the viscoelastic relaxation occurs. Second, the shearing phase is performed (keeping $z$=$z_{ini}$) at a constant velocity \unit{10}{\micro\meter\per\second} over a distance $\Delta_t$ in the following list: 11 equally spaced values between 0 and \unit{100}{\micro\meter} and 10 equally spaced values between 200 and \unit{1100}{\micro\meter}. The various shearings are performed in pseudo-random order to avoid possible biases. Third, the unloading phase is performed through vertical steps (\unit{0.5}{\micro\meter} amplitude, \unit{1}{\second} duration), from $z_{ini}$ down to a negative altitude sufficient to separate the interface. Normal and tangential forces are acquired at \unit{10}{\hertz}, so that each of the many vertical steps during the unloading phase is recorded with 10 points.

\subsection{Contact imaging}
Images of the real contact interface are taken as described in~\cite{sahli_evolution_2018}, using a high-resolution camera (Teledyne DALSA Genie Nano-GigE combined with a Qioptics optem fusion objective, \unit{2.36}{\micro\meter}/pixel): at a frame rate of 10 fps during shearing, and a single image at the end of each step during unloading. From the well-contrasted images (typical examples in Fig.~\ref{fig1}), the contact area, $A$, can be measured by segmentation using simple thresholding. When the contact is unsheared ($\Delta_t$=0), the contact is at all times circular, and the contact radius, $a$, can be defined.

\subsection{Material and interfacial properties}\label{sec:prop}

The Young's modulus $E$ of the elastomer, and the adhesion energy $w_0$ and Tabor's parameter $\lambda$ of the interface are provided in Tab.~\ref{tab1}. $\lambda$ is defined in, e.g.~\cite{maugis_adhesion_1992, maugis_contact_2000,barthel_adhesive_2008}, and quantifies the transition between the JKR (large $\lambda$, typically $>$5) and DMT (small $\lambda$, typically $<$0.1) limit models of adhesion. For each tribological pair (standard- or soft-PDMS against glass or PMMA), $E$, $w_0$ and $\lambda$ are obtained by fitting Maugis' model~\citep{maugis_adhesion_1992} on the unloading data for $\Delta_t$=0. In pratice, we fit ${a^{3/2}}/{\sqrt{6 \pi R^2}}$ as a function of ${P}/{\sqrt{6\pi a^3}}$ (as in e.g.~\cite{chaudhury_adhesive_1996,waters_mode-mixity-dependent_2010,acito_use_2023}), where $a$ is the measured contact radius. We use for those fits all available data such that $a$$<$\unit{748}{\micro\meter} for standard-PDMS/glass contacts, and such that $a$$<$\unit{936}{\micro\meter} for contacts involving soft-PDMS.

\begin{table*}[hbt!]
\centering
\begin{tabular}{p{0.4\textwidth}c|ccc}
Fitting model & Parameter & Standard/glass & Soft/glass & Soft/PMMA \\
\hline
1. \cite{maugis_adhesion_1992} on $A(P)$ & $E$ (\unit{\mega\pascal}) & 1.49$\pm$0.01 & 0.58$\pm$0.01 & 0.64$\pm$0.01 \\
(discharge with no preliminary shear) & $w_0$ (\unit{\milli\joule\per\meter^2}) & 141$\pm$2 & 252$\pm$5 & 215$\pm$2 \\
 & $\lambda$ & 1.45$\pm$0.03 & 1.23$\pm$0.04 & 1.37$\pm$0.02 \\
 \hline
2. \cite{papangelo_effect_2020} on $A(Q)$ (during preliminary shear) & $\alpha_{Pap}$ & 0.006$\pm$0.002 & 0.011$\pm$0.002 & 0.011$\pm$0.002 \\
\hline
3. Linear on $A(Q)$ (during unloading after full sliding, at $\Delta_t$=\unit{1}{\milli\meter}) & $\sigma_{unload}$ (\unit{\mega\pascal}) & 0.281$\pm$0.001 & 0.203$\pm$0.001 & 0.195$\pm$0.001\\
\hline
4. \cite{peng_effect_2021} on $P_{PO}$ (pull-off force after full sliding) & $\alpha_{Peng}$ & 0.97$\pm$0.16 & 1.00$\pm$0.14 & 1.00$\pm$0.18\\
\hline
\end{tabular}
\caption{Material and interfacial parameters}\label{tab1}
\end{table*}

For completeness of the methods, we finish by describing how we determined the values of additional interfacial parameters ($\alpha$ and $\sigma$), that will be useful only in section~\ref{sec:model}, where we perform a quantitative comparison between our experimental observations and a \rev{tentative model}. The shear index, $\alpha$, is estimated using two different models \rev{and by following the procedure proposed by their respective authors}. First, the model by \cite{papangelo_effect_2020} is fitted on the $A(Q)$ evolution during the shearing phase. $\alpha$ ($\alpha_{Pap}$ in Tab.~\ref{tab1}) and the initial contact area (before shearing) are fitted, while $E$ and $w_0$ are fixed to the values in Tab.~\ref{tab1}. In practice, for each interface and each $P_{ini}$, we use the data for $\Delta_t$=\unit{1.1}{\milli\meter} at all times before $Q$ reaches its maximum. The values (error bars) in Tab.~\ref{tab1} are the mean (standard deviation) over the three $P_{ini}$.

Second, the model by \cite{peng_effect_2021} is fitted on the value of the pull-off force obtained during the unloading phase after a preliminary shear sufficient to reach full sliding. $\alpha$ ($\alpha_{Peng}$ in Tab.~\ref{tab1}) is fitted while $E$, $w_0$, $\lambda$ and the friction strength $\sigma$ are fixed to the values in Tab.~\ref{tab1}. For each interface, we use all pull-off forces obtained for the three $P_{ini}$ and $\Delta_t$$>$\unit{90}{\micro\meter} (to ensure full sliding before separation, see Section~\ref{sec:origin}). The value of $\sigma$ ($\sigma_{unload}$ in Tab.~\ref{tab1}) is estimated beforehands by fitting a linear relationship on the $Q(A)$ evolution during the unloading phase of the experiments at $\Delta_t$=\unit{1}{\milli\meter} and for all three $P_{ini}$, each limited to the range $Q<0.8 \times max(Q)$.

In Tab.~\ref{tab1}, the error bars on $\alpha_{Pap}$  are standard deviations on the three values derived for the three $P_{ini}$. All other error bars are half of the 95\% confidence error of the fitted parameters.

\section{Results and discussion}\label{sec:resanddisc}
Typical contacts in various conditions are shown in Fig.~\ref{fig1}. The contact is circular at the end of the loading phase (Maximum indentation). At the end of the shearing phase (Maximum shear), the contact has undergone an anisotropic change in morphology: its total area is smaller and the eccentricity of its elliptic-like shape is larger when the shear displacement $\Delta_t$ is larger. Note that the above observations, although fully consistent with those of~\cite{sahli_evolution_2018, sahli_shear-induced_2019,mergel_continuum_2019}, remain incompletely understood~\citep{papangelo_shear-induced_2019, lengiewicz_finite_2020, mergel_contact_2021}. During the unloading phase, the contact progressively shrinks (keeping a non-circular shape), until full separation (Fig.~\ref{fig1}, last column).

\subsection{The pull-off force decreases with shear displacement}

\begin{figure}[b!]
\centering
\includegraphics[width=.49\linewidth]{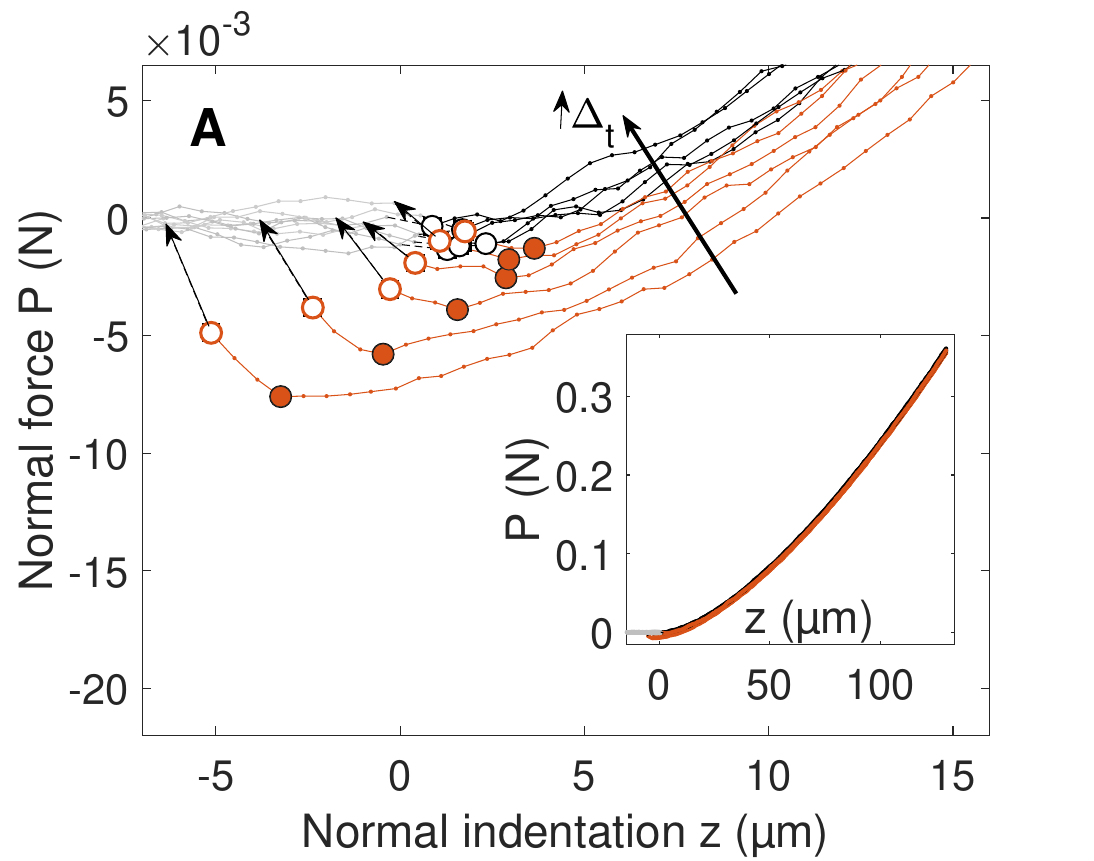}
\includegraphics[width=.49\linewidth]{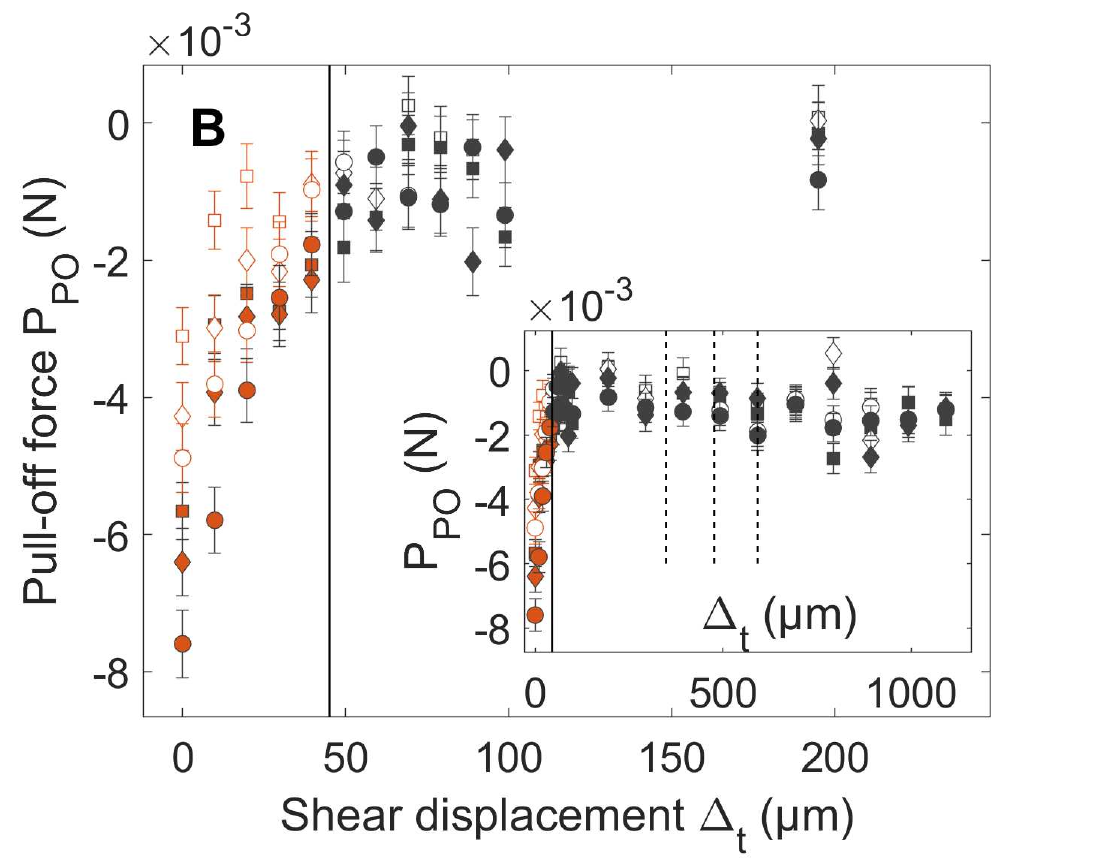}
\caption{The contact unloading path is a function of preliminary shear. Standard-PDMS/glass contacts. (A) Normal force, $P$ (average over the 10 points of each step), as a function of normal indentation, $z$, for various preliminary shear displacement $\Delta_t$ from 0 to \unit{100}{\micro\meter} (thick arrow). $P_{ini}\simeq$0.4\,N. Inset: full dataset. Main: zoom on small $P$. Solid (resp. open) disks indicate where the pull-off force $P_{PO}$ (resp. separation force $P_{Sep}$) is measured. Thin arrows point to the first data after contact separation. (B) $P_{PO}$ (solid symbols) and $P_{Sep}$ (open symbols) as functions of $\Delta_t$. Main: zoom on small $\Delta_t$. Inset: full dataset. Squares/diamonds/disks are for $P_{ini}\simeq$0.1/0.2/0.4\unit{}{\newton}. Error bars: standard deviation of the difference between the raw force signal during unloading and its trend (smoothed signal using a moving interval of $\pm$5 points). The solid vertical line indicates $\Delta_{t,c}$ (see text). The three dashed vertical lines indicate the shear displacement $\Delta_{t,s}$ at which, during the shearing phase, the friction peak is reached (from left to right: for $P_{ini}$$\simeq$0.1, 0.2 and 0.4, respectively). The color code used in both panels A and B is defined from the data in Fig.~\ref{fig3}A.}
\label{fig2}
\end{figure}

Figure~\ref{fig2}A shows the evolution of the normal force, $P$, as a function of the imposed normal displacement, $z$ ($z$=0 corresponds to first contact during loading, thus $z$ is also the normal indentation)\rev{, on the example of a standard-PDMS/glass contact}. Each curve corresponds to a specific value of the initial shear displacement, $\Delta_t$. All curves are found different, indicating that the shearing phase affects the subsequent unloading path of the contact. From each curve in Fig.~\ref{fig2}A, one can measure two characteristic forces: the minimum value of $P$, so-called pull-off force, $P_{PO}$ (solid disk) ; the force of the last step spent in contact (as verified on contact images), defining the separation force, $P_{Sep}$ (open disk). The curve for $\Delta_t$=0 is the one that would have been obtained in a classical pull-off test (no preliminary shear applied). Note that $\left| P_{PO}\right|>\left| P_{Sep}\right|$, as expected in pull-off tests under controlled normal displacement~\citep{maugis_contact_2000}.

The evolution of $P_{PO}$ with $\Delta_t$ (Fig.~\ref{fig2}B, solid symbols) indicates the existence of two regimes. First, for increasing values of $\Delta_t$ from $\Delta_t$=0 up to a critical value $\Delta_{t,c}\simeq$~\unit{40-50}{\micro\meter}, the amplitude of $P_{PO}$ steadily decreases, starting from the (negative) value one would find in a classical pull-off test in absence of preliminary shearing. Second, for values of $\Delta_t$ larger than  $\Delta_{t,c}$, $P_{PO}$ becomes independent of $\Delta_t$, and takes a non-vanishing residual value, with a mean value about 12\% of its initial value. The crossover between the two regimes is rather sharp. The open symbols in Fig.~\ref{fig2}B show that all those qualitative observations also apply to $P_{Sep}$.

The results of Fig.~\ref{fig2}B validate our initial hypothesis that the pull-off force is strongly affected by minute preliminary shear. They fill the gap between two known results: (i) the classical pull-off force is recovered when $\Delta_t$=0 and (ii) the pull-off force is much reduced, although non-vanishing, when $\Delta_t$ is so large that the contact is fully sliding before unloading starts~\citep{peng_effect_2021}. The latter case corresponds to the data points on the right of the vertical dashed lines in Fig.~\ref{fig2}B ($\Delta_t$ beyond a few~\unit{100}{\micro\meter}), for which full sliding is observed during the shearing phase (after a peak in the tangential force). Importantly, our results shed light on a characteristic shear displacement, $\Delta_{t,c}$, of order a few~\unit{10}{\micro\meter}, already sufficient to lower adhesive forces down to their minimum. In practice, shear displacements as small as \unit{10}{\micro\meter} (i.e. $\sim$100 times smaller than the contact radius and $\sim$35-60 times smaller than the shear displacement necessary to trigger full sliding during the shearing phase, $\Delta_{t,s}$) already reduce the pull-off force by $\sim$25\% (Fig.~\ref{fig2}B). Such a rapid decrease sets strong constraints on the required orthogonality between interface plane and pulling direction in adhesion setups, in order to avoid any unwanted tangential force during loading/unloading, that would induce biases in the measurement of adhesion forces.

\subsection{Origin of the transition at $\Delta_{t,c}$}\label{sec:origin}

\begin{figure}[t!]
\centering
\includegraphics[width=.49\linewidth]{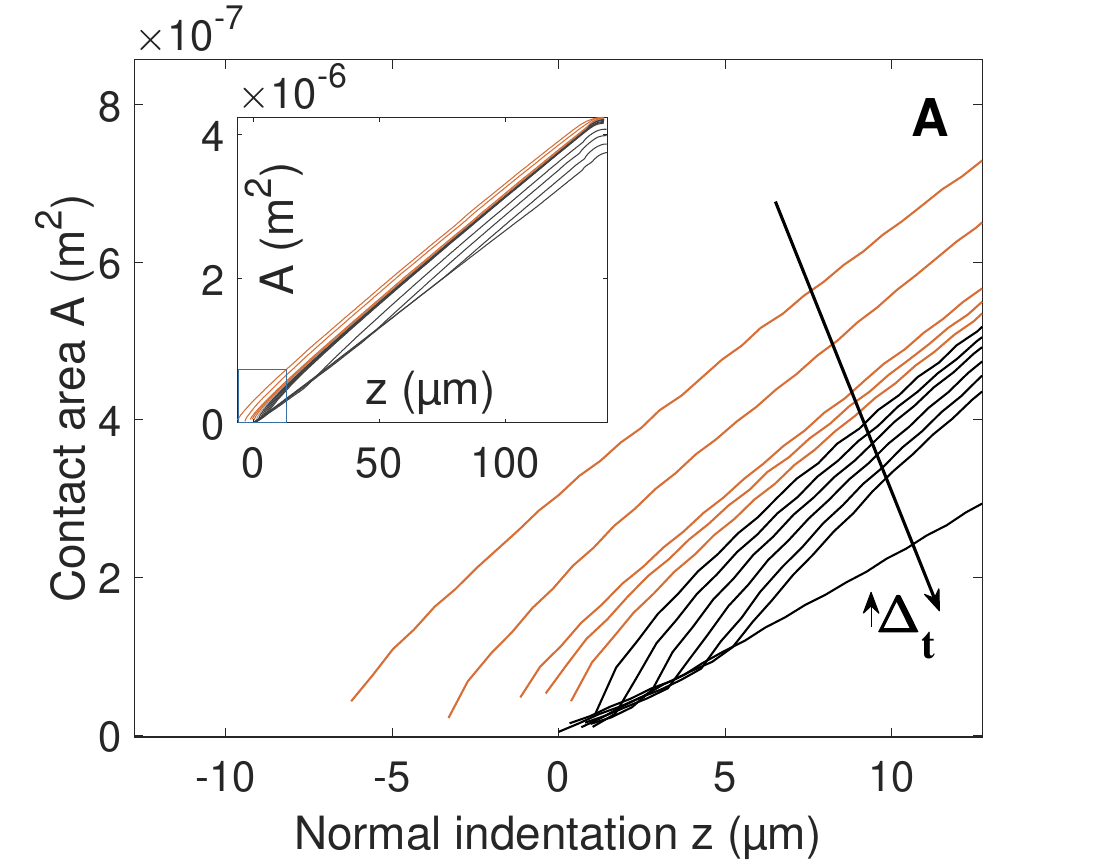}
\includegraphics[width=.49\linewidth]{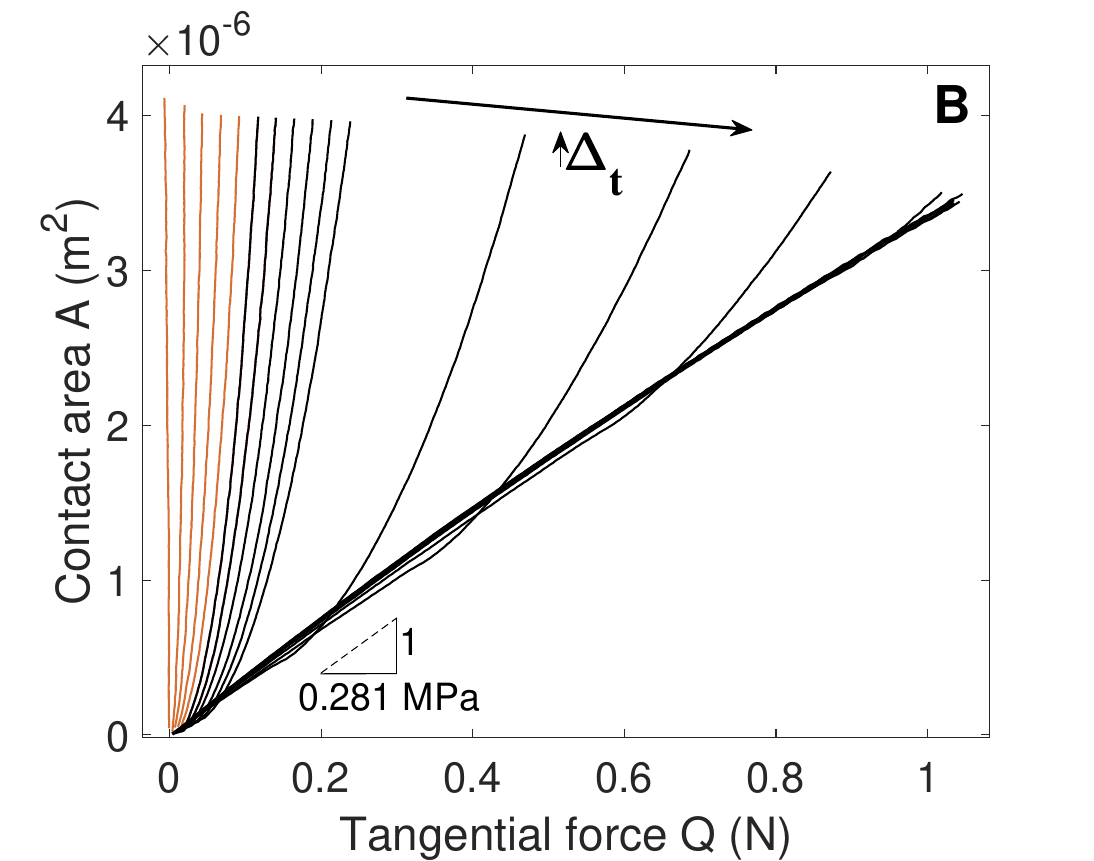}
\caption{Evolution of the contact area during unloading. Contact area, $A$, as a function of either (A) the normal indentation $z$ or (B) the tangential force $Q$, during the unloading phase of a standard-PDMS/glass contact. Different curves are for different $\Delta_t$. $P_{ini}\simeq$0.4\,N. (A) For clarity, only the curves for the 11 smallest and the largest $\Delta_t$ are shown in the main plot (zoom on small $z$). Inset: full data, with the zoomed region indicated as a blue rectangle. Curves not showing an abrupt change in slope \rev{as the contact area decreases} are highlighted in red. This color code is also used in B and in Fig.~\ref{fig2}. (B) \rev{Also in this plot, the red curves end at small contact areas without reaching the linear master curve on which all black curves end.} The dashed line represents the fitted slope of the linear master curve (see corresponding value of $\sigma_{unload}$ in Tab.~\ref{tab1}), which is reached when the contact is fully sliding.}
\label{fig3}
\end{figure}

To pinpoint the differences in the unloading of contacts having undergone preliminary shear either below or above $\Delta_{t,c}$, we harness the additional information contained in the contact images, and in particular the evolution of the contact area $A$. Figure~\ref{fig3}A shows how $A$ decreases as the indentation $z$ decreases during unloading, for various values of the preliminary shear distance. For the smallest $\Delta_t$, the curves reach their minimum area (at separation) through a single nonlinear branch (cases highlighted in red). In contrast, for larger $\Delta_t$, the curves  are made of two different branches: an initial nonlinear branch for large $A$, analogous to that at small $\Delta_t$ ; a second linear-like branch for small $A$. We emphasize that all such second branches tend to collapse on the same master curve, suggesting that, in this range of $\Delta_t$, all contacts reach a common state before separation, irrespective of the amount of preliminary shear. Most importantly, the curves made of a single branch correspond exactly to those for which the pull-off force is in the first, $\Delta_t$-dependent, regime ($\Delta_t<\Delta_{t,c}$), as visible when applying the colorcode of Fig.~\ref{fig3}A to Fig.~\ref{fig2}B. Thus, we conclude that the transition between the two adhesion regimes at $\Delta_{t,c}$, as seen in Fig.~\ref{fig2}B, has the very same physical origin as the emergence of a second, linear branch in the curves $A(z)$ of Fig.~\ref{fig3}A.

To interpret the nature of the common branch on which all $A(z)$ curves in the second adhesion regime end up, we consider the concurrent evolution of the contact area $A$ and of the tangential force $Q$ (Fig.~\ref{fig3}B). For each curve, both $A$ and $Q$ decrease during unloading and monotonously approach the origin. The initial point (topmost point of each curve) is the point reached at the end of the shearing phase. The initial area decreases when $\Delta_t$ increases and, for each $P_{ini}$, the ensemble of all initial points delineate a quadratic-like decrease of the curve $A(Q)$, in agreement with previous studies on sheared elastomer contacts using the same PDMS as that used here~\citep{waters_mode-mixity-dependent_2010,sahli_evolution_2018,sahli_shear-induced_2019,mergel_continuum_2019}. Such contacts are known to obey a friction law such that, when the contact is fully sliding, $A$ is proportional to $Q$, the proportionality coefficient being the friction strength of the interface, $\sigma$~\citep{cohen_incidence_2011,sahli_evolution_2018,mergel_continuum_2019,lengiewicz_finite_2020}. In Fig.~\ref{fig3}B, all $A(Q)$ curves marked in black (second adhesion regime) are found to end up (sooner for larger $\Delta_t$) on a linear master curve reminiscent of the above-mentioned friction law, suggesting that the master curve actually indicates full sliding of the contact. This interpretation is strongly grounded by the value of the inverse slope of the master curve, about \unit{0.28}{\mega\pascal} (Fig.~\ref{fig3}B and Tab.~\ref{tab1}), very close to the values of $\sigma$ found for similar standard-PDMS/glass interfaces in~\cite{sahli_evolution_2018, mergel_continuum_2019, lengiewicz_finite_2020}.

Thus, the scenario that emerges from our data is that, in the shear-dependent adhesion regime ($\Delta_t<\Delta_{t,c}$, red curves), the preliminary shear is too small to yield any full sliding of the contact before the two solids separate. In contrast, in the shear-independent adhesion regime ($\Delta_t>\Delta_{t,c}$, black curves), the initial shear is large enough to make the contact enter, at some point during unloading, into a fully sliding state, during which $A$ and $Q$ are proportional. This is consistent with the extremely non-circular shape of the contact at separation (see the two bottom-most images in the right-most column of Fig.~\ref{fig1}), which is expected for a fully sliding contact. We emphasize that full sliding in our experiments can be reached either already during the shearing phase or only during the unloading phase. When $\Delta_t$ is larger than the shear displacement required to trigger full sliding during the shearing phase, $\Delta_{t,s}$ ($\simeq$ 350/475/600\unit{}{\micro\meter}, i.e. about 8/11/14 times $\Delta_{t,c}$ for standard-PDMS/glass contacts at $P_{ini}\simeq$~0.1/0.2/0.4\unit{}{\newton}, see dashed vertical lines in Fig.~\ref{fig2}B), our experimental conditions are similar to those used by~\cite{peng_effect_2021}, where unloading was always performed after the contact has undergone a significant amount of steady macroscopic sliding during shearing. In those conditions, it is expected that the pull-off force is independent of $\Delta_t$ because, at the end of the shearing phase, the contact state is identical for all $\Delta_t$ at the end of the shearing phase and the contact spends all of the ensuing unloading phase in full sliding conditions (see Fig.~\ref{fig3}B). In contrast, when $\Delta_{t,c}$$<$$\Delta_t$$<$$\Delta_{t,s}$, the contact exits the shearing phase without having reached gross sliding. Nevertheless, full sliding is triggered during the subsequent unloading phase, due to the fact that the maximum shear displacement that a contact can sustain without sliding is smaller for a smaller contact. Because the macroscopic shear displacement $\Delta_t$ is kept constant during unloading and, simultaneously, the contact is continuously shrinking, the interface eventually reaches its sliding threshold. Note that, consistently with~\cite{lengiewicz_finite_2020,prevost_probing_2013}, full sliding is expected to be preceded by a so-called partial slip state, where a peripheral region of the contact is slipping while the central region has remained stuck. In summary, for any value of $\Delta_t$ larger than $\Delta_{t,c}$, the contact state at separation is expected to be a fully sliding one, irrespective of when full sliding started (during either the shearing or unloading phase), and the pull-off force is a constant (points on the right of the solid vertical line in Fig.~\ref{fig2}B).

\subsection{Robustness of the observations}
\begin{figure}[hbt!]
\centering
\includegraphics[width=0.49\textwidth]{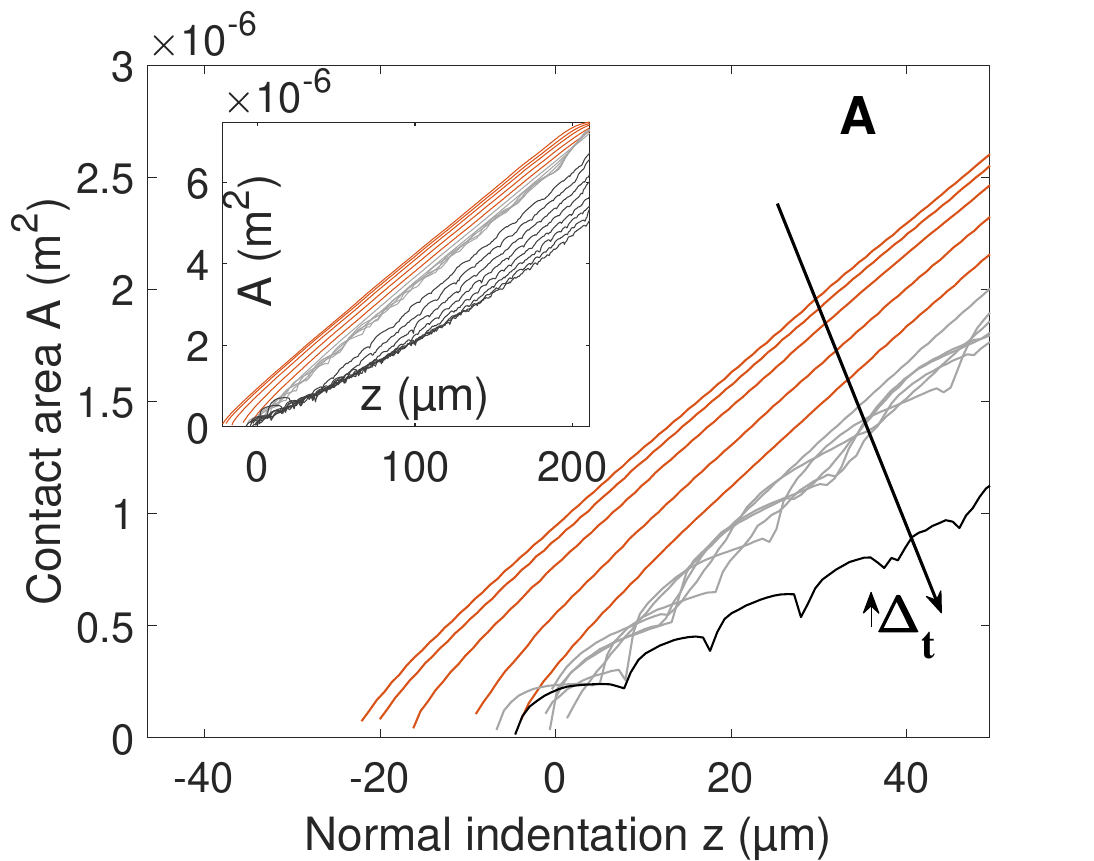}
\includegraphics[width=0.49\textwidth]{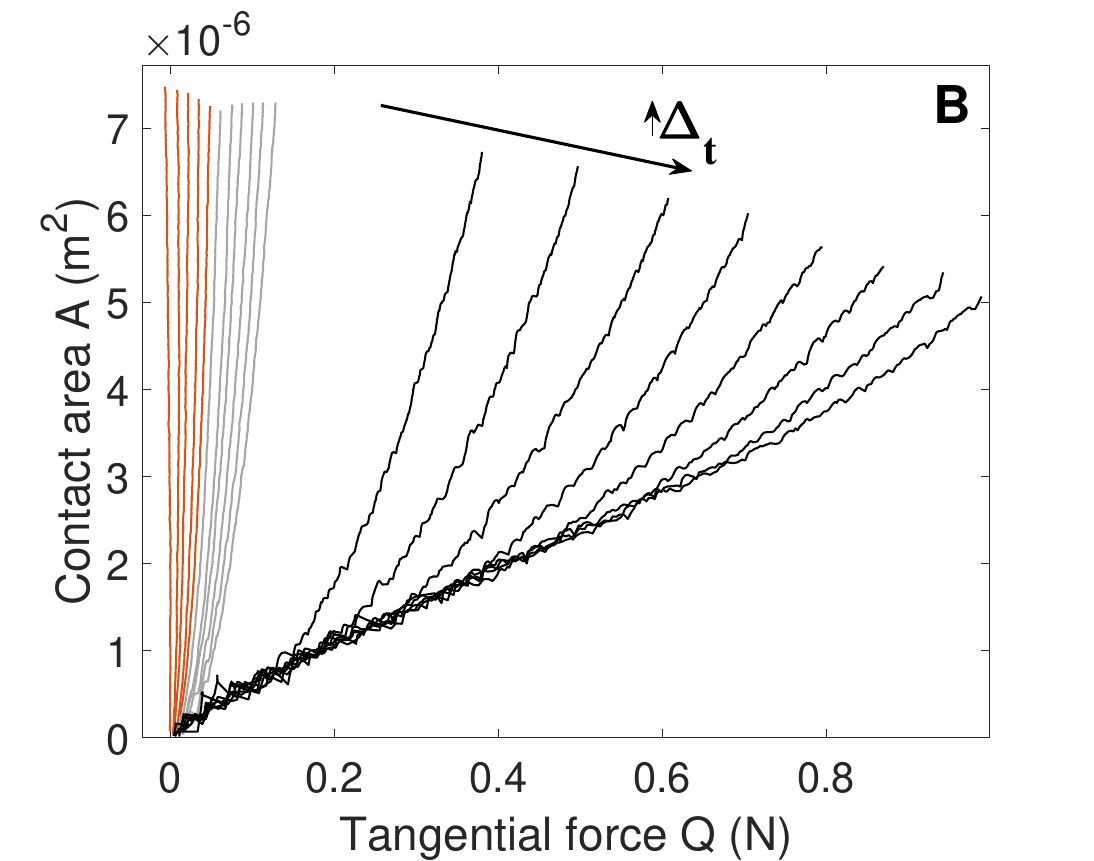}
\includegraphics[width=0.49\textwidth]{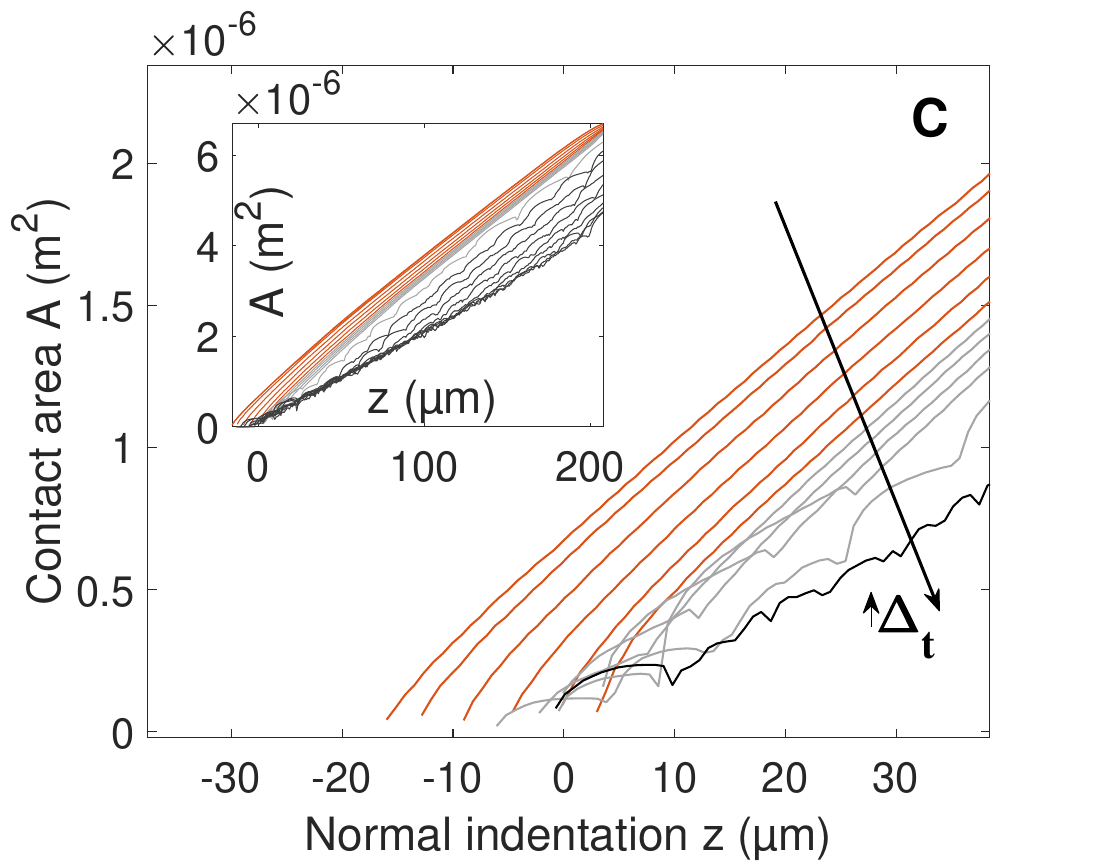}
\includegraphics[width=0.49\textwidth]{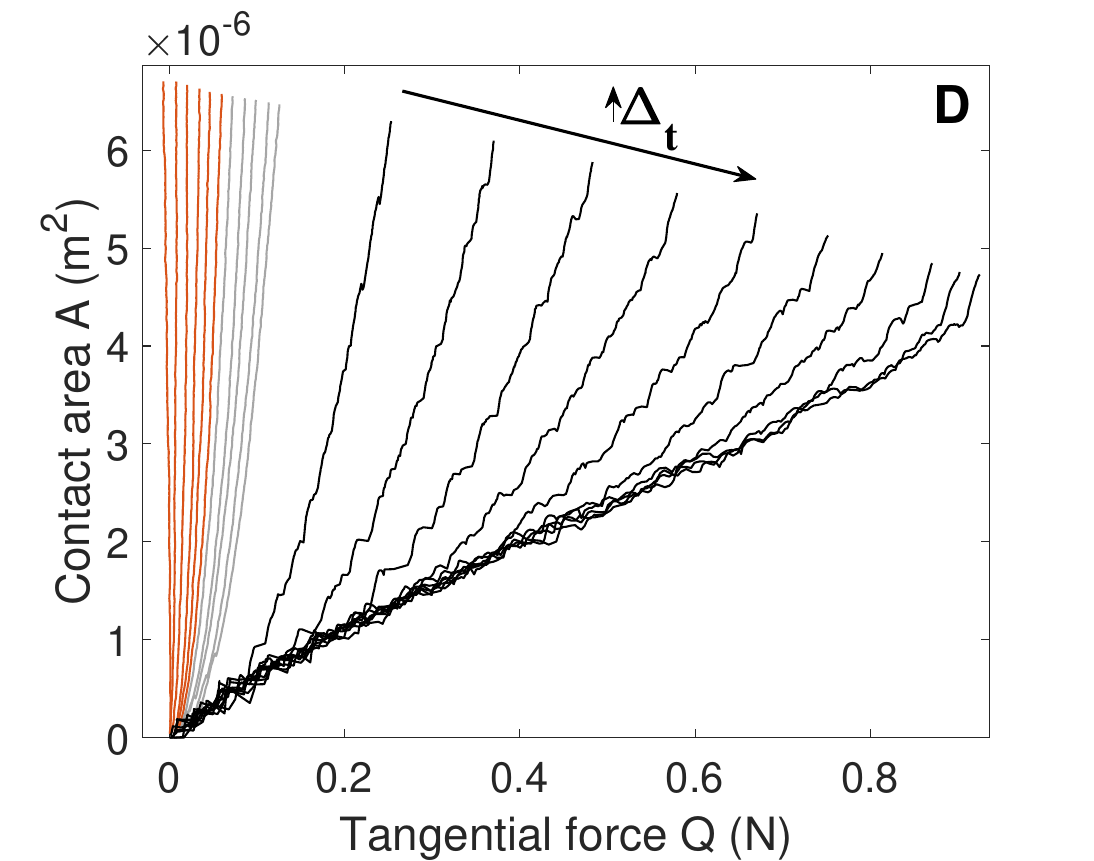}
\caption{Evolution of the contact area during unloading of interfaces involving soft-PDMS. Contact area, $A$, as a function of the normal indentation $z$ (left column) or of the tangential force $Q$ (right column). Top line (resp. bottom line): soft-PDMS/glass (resp. soft-PDMS/PMMA). \rev{$P_{ini} \simeq$~\unit{0.4}{\newton}.} For clarity, the main plots of panels A and C only show the curves for the 11 smallest and the largest $\Delta_t$. The wiggling curves in panels A and C illustrate that contact instabilities (see text) decorate the area evolution of interfaces involving soft-PDMS, and significantly increase the uncertainty in the determination of $\Delta_{t,c}$ from $A(z)$ curves. \rev{The color code (red and black curves) is analogous to that defined in Fig.~\ref{fig3}, and used in Figs.~\ref{fig2} and~\ref{fig3}.} Grey curves in A and C correspond to cases where one cannot firmly conclude about whether the $A(z)$ curves are made of a single branch (red curves) or two branches (black curves) \rev{due to contact instabilities (see text)}.}\label{fig4}
\end{figure}

\begin{figure}[hbt!]
\centering
\includegraphics[width=0.49\textwidth]{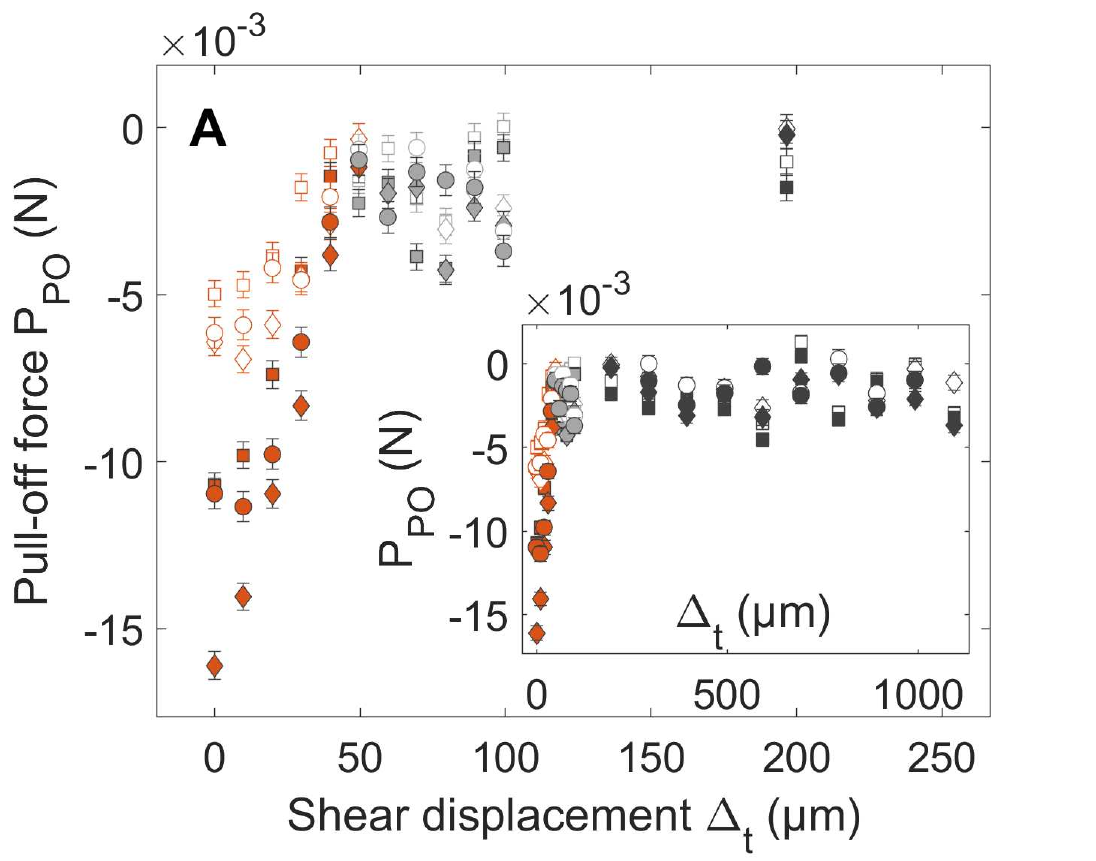}
\includegraphics[width=0.49\textwidth]{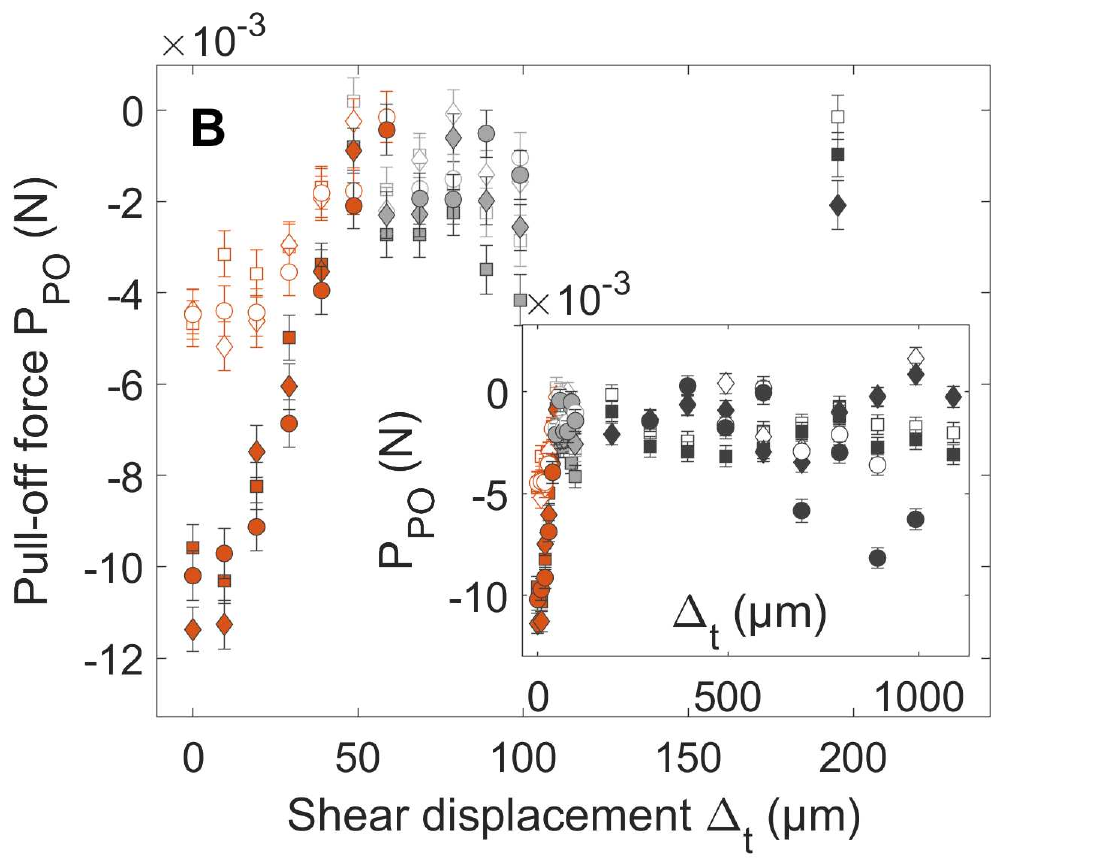}
\caption{Evolution of the pull-off force $P_{PO}$ (solid markers) and of the separation force $P_{sep}$ (open markers) as a function of $\Delta_{t}$, for the experiments involving soft-PDMS. (A) Soft-PDMS/glass interface. (B) Soft-PDMS/PMMA interface. Insets: full datasets. Mains: zooms on small $\Delta_{t}$. Squares/diamonds/disks are for $P_{ini}$$\simeq$ 0.1/0.2/0.4\unit{}{\newton}. Error bars: standard deviation of the difference between the raw force signal and its trend (smoothed signal using a moving interval of $\pm$5 points). Color code similar to that of Figs.~\ref{fig2}, \rev{\ref{fig3} and ~\ref{fig4}}. Red (resp. black): points corresponding to $A(z)$ curves with a single branch (resp. with a transition between two branches). Grey: points for which the attribution of a red or black color was dubious, due to the oscillations in the $A(z)$ curve, as seen in Figs.~\ref{fig4}A and~\ref{fig4}C \rev{(grey points in the present figure correspond to the grey curves in Fig.~\ref{fig4})}.}\label{fig5}
\end{figure}

\begin{figure}[hbt!]
\centering
\includegraphics[width=0.33\textwidth]{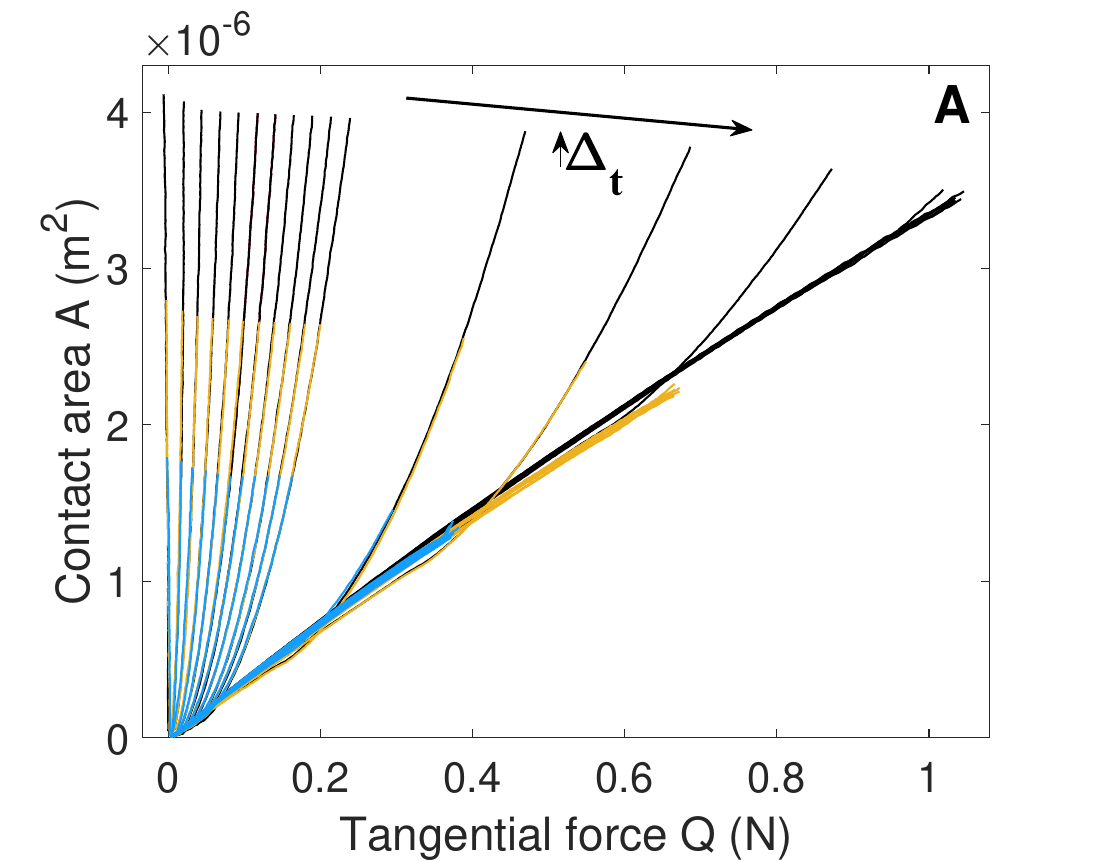}
\includegraphics[width=0.33\textwidth]{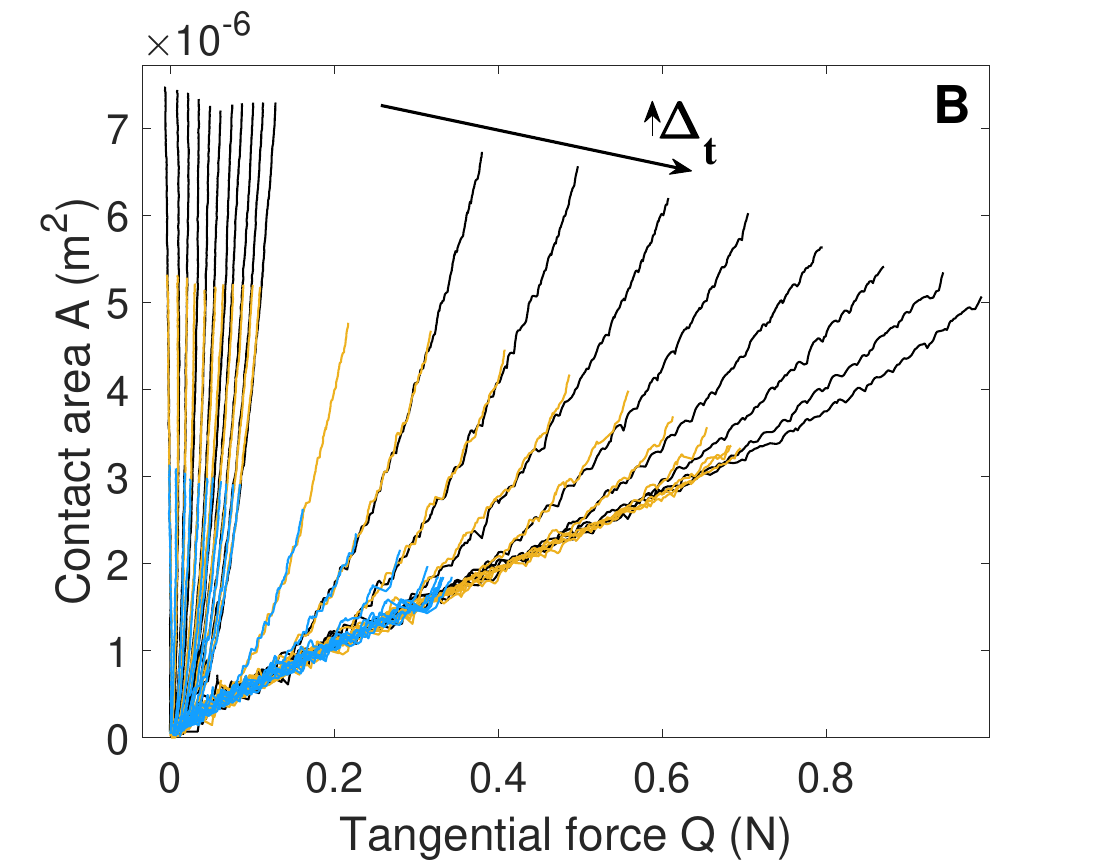}
\includegraphics[width=0.33\textwidth]{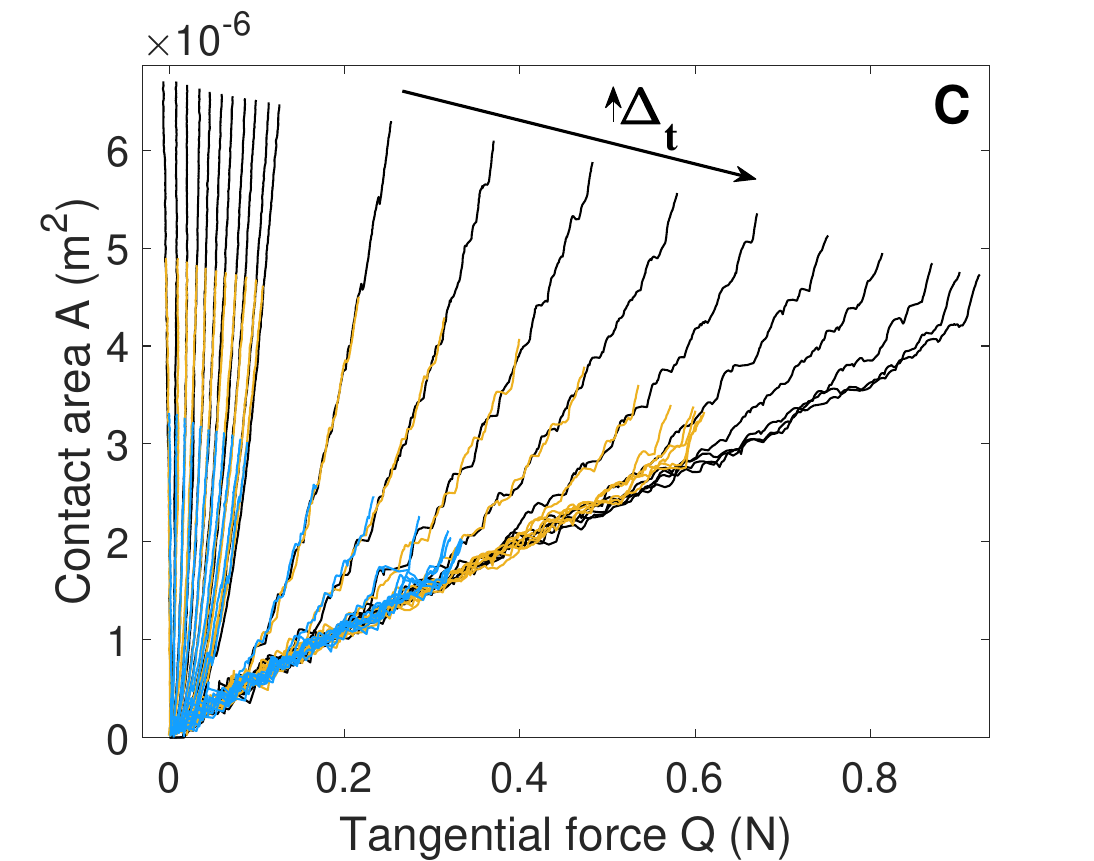}
\caption{\rev{The results are robust against changes in initial normal load. Contact area, $A$, as a function of the tangential force, $Q$, during the
unloading phase, for the three types of interface. (A) Standard-PDMS/glass. (B) Soft-PDMS/glass. (C) Soft-PDMS/PMMA. In each panel, the three colors correspond to three different $P_{ini}$: blue for $P_{ini}\simeq$\unit{0.1}{\newton}, yellow for $P_{ini}\simeq$\unit{0.2}{\newton} and black for $P_{ini}\simeq$\unit{0.4}{\newton}. The black curves in panel A are the same as the curves in Fig.~\ref{fig3}B. Note how well do the curves for the same $\Delta_t$ but different $P_{ini}$ superimpose.}}\label{fig6}
\end{figure}

To check whether the results shown in Figs.~\ref{fig2} and~\ref{fig3} on standard-PDMS/glass contacts submitted to $P_{ini}\simeq\unit{0.4}{\newton}$ are robust to changes made on the tribological system, we performed complementary experiments, changing the properties of the interface and the initial normal load (see \rev{Figs.~\ref{fig4} to~\ref{fig6}}). 

First, the very same loading-shearing-unloading experimental protocol was reproduced with two different types of interfaces (see section~\ref{sec:sample}): a softer PDMS against either a glass or a PolyMethylMethAcrylate (PMMA) plate. As shown in Figs.~\ref{fig4} and~\ref{fig5}, all qualitative features observed in Figs.~\ref{fig2} and~\ref{fig3} are robust to those material changes. The only difference is the occurrence, during unloading, of contact instabilities (presumably reattachment folds, common in the literature on sheared elastomers, see~\cite{petitet_materiaux_2008, lengiewicz_finite_2020}) that decorate the area evolution and significantly increase the uncertainty in the determination of $\Delta_{t,c}$ from $A(z)$ curves. Note however that the transition from a decreasing to a constant pull-off force as a function of $\Delta_t$ (Fig.~\ref{fig5}) apparently still occurs at about~\unit{50}{\micro\meter}, suggesting that $\Delta_{t,c}$ is actually only weakly affected by our changes in contacting materials.

Second, we varied the initial normal load (to $\simeq$~0.2 and \unit{0.1}{\newton}) and found that the results remain essentially identical. \rev{This is illustrated on the example of the curves $A(Q)$ in Fig.~\ref{fig6}, where the data for the three tested $P_{ini}$ are overplotted, for each of the three different types of interfaces}. Although curves corresponding to larger initial indentations start with larger forces and areas, they do overlap with those corresponding to smaller initial indentations, in their common range. Such an observation indicates that the final path to interface separation is independent of the initial compressive state (and so $\Delta_{t,c}$ is independent of $P_{ini}$). Thus, for our specific loading protocol, the contact state during the unloading phase is presumably controlled solely by the instantaneous couple of imposed displacements: $z$ and $\Delta_t$. This is consistent with the data shown in Figs.~\ref{fig2}B and~\ref{fig5}, which indicate that the evolutions of $P_{PO}$ and $P_{Sep}$ are unaffected by $P_{ini}$.

\subsection{Modelling attempt}\label{sec:model}
To capture our experimental results, one needs a model describing adhesion in mixed-mode conditions (normal + tangential forces). Various such models exist, with a recent interest in explaining anisotropic shear-induced contact shrinking observed in \cite{sahli_evolution_2018, mergel_continuum_2019, sahli_shear-induced_2019}. While investigating the relevance of numerical models coupling adhesion and friction~\citep{mergel_continuum_2019, mergel_contact_2021, perez-rafols_incipient_2022} to our sheared adhesion tests is left for future work, here, in an effort to unravel the physical parameters that control $\Delta_{t,c}$, we consider  analytical models such as those of \cite{papangelo_mixed-mode_2019, papangelo_shear-induced_2019, das_sliding_2020, papangelo_effect_2020, peng_effect_2021}. Most such analytical mixed-mode adhesion models assume axisymmetric contacts, which is less and less realistic when the contact is increasingly sheared up to full sliding (Fig.~\ref{fig1}). To our knowledge, the only analytical model describing non-circular contacts is in~\cite{papangelo_shear-induced_2019}. Unfortunately, it is valid in the limit of short-ranged adhesion, i.e. that of the classical JKR adhesion model~\citep{johnson_surface_1971}. In contrast, and consistently with~\cite{acito_use_2023} on PDMS, our contacts have a Maugis parameter $\lambda$ between 1 and 2 (Tab.~\ref{tab1}), far from the JKR limit ($\lambda$$>$5).

In the absence of a fully suitable analytical model, we propose a first, simple modelling attempt limited to qualitative comparisons in the JKR limit. We consider the axisymmetric model of \cite{papangelo_effect_2020} and test if it predicts our main result, i.e. a transition between two pull-off regimes at a critical $\Delta_t$.
The model describes the adhesive contact (nominal work of adhesion $w_0$) between a linear elastic sphere (radius $R$, reduced Young modulus $E^*$) and a rigid plate. A fracture description of the (circular) contact's periphery is made, with a shear-dependent energy release rate, $G_{eff}=w_0-\alpha \frac{E^* \Delta_t^2}{3 \pi a}$, $a$ being the contact radius. $\alpha \in [0;1]$ is a dimensionless index (used e.g. in~\cite{mcmeeking_interaction_2020,ciavarella_degree_2020,peng_effect_2021}) quantifying the part of the mode II (shear mode) energy release rate that is dissipated at the interface (if $\alpha$=0, there is no reversible slip, so that slip does not affect adhesion).

The constitutive equation of the model, which relates the contact radius, $a$, to the couple of imposed displacements, $\Delta_n$ and $\Delta_t$, is (obtained by combining Eqs.~(19) and~(20) of~\cite{papangelo_effect_2020}:
\begin{equation}
\rev{\Delta_n=\frac{a^2}{3R}+\left(1+\frac{2E^*a}{k_n}\right)\left[\frac{2a^2}{3R}-\sqrt{\frac{2 \pi a w_0}{E^*}-\frac{2}{3}\alpha \left(\frac{\Delta_t}{1+\frac{4 E^* a}{3 k_t}}\right)^2}\right],}\label{Eq:ConstFull}
\end{equation}
\rev{where $k_n$ and $k_t$ are the loading apparatus stiffnesses, in the normal and tangential directions, respectively. In our mechanical device, $k_n$=2.8$\times$10$^{5}$\unit{}{\newton\meter} and $k_t$=1.2$\times$10$^{5}$\unit{}{\newton\meter}. These values are so large that, in all our experiments, $\frac{2E^*a}{k_n}$ and $\frac{4E^*a}{3k_t}$ remain always smaller than 0.016 and 0.025 , respectively. We further observe that the contact area at separation is always smaller than \unit{0.1}{\milli\meter\squared} (for standard-PDMS) and \unit{0.2}{\milli\meter\squared} (for soft-PDMS), as seen in Figs.~\ref{fig3}A, \ref{fig4}A and \ref{fig4}B. With those values, $\frac{2E^*a}{k_n}$ and $\frac{4E^*a}{3k_t}$ are actually always smaller than 0.0039 and 0.0025, respectively, when the two surfaces detach, which is the phenomenon under study. Thus, in the following, we will neglect $\frac{2E^*a}{k_n}$ and $\frac{4E^*a}{3k_t}$, so that the constitutive equation reduces to:}

\begin{equation}
\Delta_n=\frac{a^2}{R}-\sqrt{\frac{2 \pi a w_0}{E^*}-\frac{2}{3}\alpha \Delta_t^2}.\label{Eq:Const}
\end{equation}

\rev{To mimic the unloading phase of our experiments, we apply this model with fixed $\Delta_t$ and imposed decreasing normal indentation $\Delta_n$ (equivalent to $z$ in Figs.~\ref{fig1}--\ref{fig3} and~\ref{fig4}). The decrease of $\Delta_n$ causes a decrease of the contact radius $a$, but Eq.~\ref{Eq:Const} cannot describe the contact down to a vanishing $a$, because two phenomena may occur beforehand.}

\begin{figure}[hbt!]
\centering
\includegraphics[width=\textwidth]{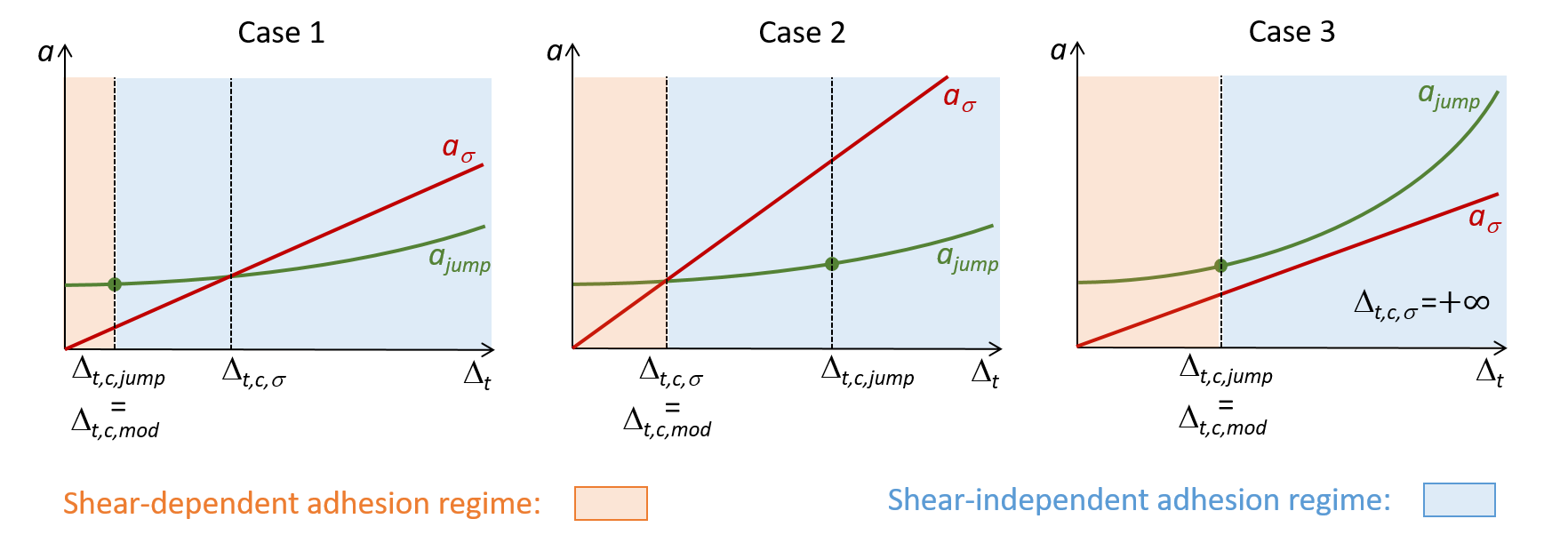}
\caption{\rev{Sketch of various possible evolutions of $a_{jump}$ (green curves, solution of Eq.~\ref{Eq:unsta}) and $a_{\sigma}$ (red straight lines, Eq.~\ref{Eq:asigma}) as functions of $\Delta_t$. Green dots: points where $\Delta_{n,jump}$=0, and which define $\Delta_{t,c,jump}$. $\Delta_{t,c,\sigma}$ is defined as the intersection of both curves, and is set to $+\infty$ when such intersection does not exist (case 3). Case 1 and 2 differ by the respective locations of $\Delta_{t,c,jump}$ and $\Delta_{t,c,\sigma}$ along the horizontal axis. $\Delta_{t,c,mod}$, defined as the minimum of $\Delta_{t,c,jump}$ and $\Delta_{t,c,\sigma}$, separates the two adhesion regimes: a shear-dependent adhesion regime at small $\Delta_t$ (orange region) and a shear-independent adhesion regime at large $\Delta_t$ (blue region).}}\label{figmodel}
\end{figure}

\rev{The first phenomenon is the so-called jump instability, described in~\cite{papangelo_mixed-mode_2019}. It is analogous to the instability that leads to contact separation in the classical JKR model. It occurs when the slope of the evolution of $a$ as a function of $\Delta_n$ becomes infinite. This takes place when the contact radius is equal to $a_{jump}$, defined as $\frac{\partial \Delta_n}{\partial a}|_{a=a_{jump}}$=0, i.e. when (from Eq.~\ref{Eq:Const}):}
\begin{equation}
\rev{\sqrt{\frac{2 \pi a_{jump} w_0}{E^*}-\frac{2}{3}\alpha \Delta_{t}^2}=\frac{\pi w_0 R}{2 a_{jump} E^*}.}\label{Eq:unsta}
\end{equation}
\rev{Solving Eq.\ref{Eq:unsta}, one obtains the evolution of $a_{jump}$ as a function of $\Delta_t$. As sketched as a green curve in Fig.~\ref{figmodel}, it is a monotonically increasing and accelerating function, starting from a finite value $\left(\frac{\pi w_0 R^2}{8E^*}\right)^{1/3}$ for $\Delta_t$=0, which corresponds to the classical JKR result for fixed-grips conditions.}

\rev{The second phenomenon possibly occurring during the unloading phase is the transition to sliding. The rapid decrease of the contact area during unloading of a sheared contact, as seen e.g. in Fig.~\ref{fig3}B, causes an increase of the average shear stress on the contact, $Q/A$. Such an increasing stress might reach the sliding friction stress of the interface, $\sigma$, thus precipitating full sliding of the contact. Such a criterion ($Q=\sigma A$) for the transition to sliding in terms of a critical interfacial shear strength has been used in many models from the literature on mixed-mode loaded soft contacts, including~\cite{mergel_continuum_2019,mergel_contact_2021,lengiewicz_finite_2020,scheibert_onset_2020}. Within the model of~\cite{papangelo_effect_2020}, $Q=\frac{4}{3}E^* a \Delta_t$ and the contact is assumed to the circular. Thus, the sliding criterion translates into a critical contact radius $a_{\sigma}$ given by:}
\begin{equation}
\rev{a_{\sigma}=\frac{4}{3 \pi}\frac{E^*}{\sigma}\Delta_t.}\label{Eq:asigma}
\end{equation}
\rev{$a_{\sigma}$ is linearly related to $\Delta_t$, as sketched as a red straight line in Fig.~\ref{figmodel}}.

\rev{We now assess for which conditions a contact described by the above modelling framework is likely to enter a fully sliding regime before contact separation.}
\begin{itemize}
\item \rev{If, for a given $\Delta_t$, $a_{\sigma}>a_{jump}$ (i.e., for $\Delta_t$ larger than that at the intersection of the red and green curves, see cases 1 and 2 in Fig.~\ref{figmodel}) then, when the contact radius $a$ decreases due to unloading, the transition to full sliding will occur before the jump instability can take place. According to our experimental observations, such a contact that slides before separation will belong to the shear-independent adhesion regime.}
\item \rev{If, conversely, $a_{jump}>a_{\sigma}$ (i.e., for $\Delta_t$ smaller than that at the intersection of the red and green curves, see case 3 in Fig.~\ref{figmodel} and cases 1 and 2), the unloading contact will undergo the jump instability before the criterion for the transition to sliding is met . At that point, one must distinguish two cases.}
\begin{itemize}
\item \rev{If the normal displacement at which the jump instability occurs is negative ($\Delta_{n,jump}<0$, green curves on the left of the green dots in Fig.~\ref{figmodel}), i.e. the contact is tensile just before instability, then the contact will abruptly separate without any prior sliding, and the contact will belong to the shear-dependent adhesion regime.}
\item \rev{If, in contrast, $\Delta_{n,jump}>0$ (green curve on the right of the green dots in Fig.~\ref{figmodel}), then the contact remains in a compressive state after instability, so that the solids cannot separate yet. Thus, the contact finishes its unloading phase following a branch different from that described by the model of Eq.~\ref{Eq:Const} (this initial branch ceases to exist for $\Delta_n<\Delta_{n,jump}$). Consistently with our observations, and with an assumption already used in~\cite{xu_asperity-based_2022}, we assume that in this other branch the contact still exists and is fully sliding. Therefore, such a contact will belong to the shear-independent adhesion regime.}
\end{itemize}
\end{itemize}

\rev{In this context, in order to determine the critical tangential displacement $\Delta_{t,c}$, one must perform two calculations. The first is to calculate the value of $\Delta_t$ for which $\Delta_{n,jump}$=0. To do this, one replaces the square root in Eq.~\ref{Eq:Const} by the right hand term of Eq.~\ref{Eq:unsta}, to get the following relationship between the indentation and contact radius at unstable separation, respectively $\Delta_{n,jump}$ and $a_{jump}$:}
\begin{equation}
\rev{\Delta_{n,jump}=\frac{a_{jump}^2}{R}-\frac{\pi w_0 R}{2 a_{jump} E^*}.}\label{Eq:critical} 
\end{equation}
\rev{Note that for $\Delta_t$=0, we recover the classical JKR value for fixed-grips conditions~\citep{maugis_contact_2000}: $\Delta_{n,jump,JKR}=-\frac{3}{4}\left(\frac{\pi^2 w_0^2 R}{{E^*}^2}\right)^{1/3}$. Let us  denote by $a_{jump,c}$ the solution of Eq.~\ref{Eq:critical} when $\Delta_{n,jump}$=0. Then, the solution of Eq.~\ref{Eq:Const} for $\Delta_{n}$=0 and $a=a_{jump,c}$ is:}
\begin{equation}
\rev{\Delta_{t,c,jump}=\frac{3}{2^{7/6}\sqrt{\alpha}}\left(\frac{\pi^2 w_0^2 R}{E^{*2}}\right)^{1/3}.}
\end{equation}
\rev{$\Delta_{t,c,jump}$ is the expected value of $\Delta_{t,c}$ if only the jump instability is possible, i.e., excluding the transition to sliding (see green dots in Fig.~\ref{figmodel}). Interestingly, $\Delta_{t,c,jump}$ and JKR's indentation at separation $\Delta_{n,jump,JKR}$ are proportional: $\Delta_{t,c,jump}=\frac{2^{5/6}}{\sqrt{\alpha}} \left|\Delta_{n,jump,JKR}\right|$. }

\rev{The second calculation consists in finding $\Delta_{t,c,\sigma}$, the value of $\Delta_t$ such that $a_{\sigma}=a_{jump}$. The equation to be solved is Eq.~\ref{Eq:unsta}, in which $a_{jump}$ is replaced by the expression of $a_{\sigma}$ in terms of $\Delta_t$ as given in Eq.~\ref{Eq:asigma}. We have not found an analytical solution for $\Delta_{t,c,\sigma}$, and thus evaluate it numerically as the intersection of $a_{\sigma}(\Delta_t)$ (Eq.~\ref{Eq:asigma}) and $a_{jump}(\Delta_t)$, as sketched in Fig.~\ref{figmodel}. Note that in certain conditions (case 3 in Fig.~\ref{figmodel}), such intersection does not exist, in particular when $\alpha$ and/or $\sigma$ are sufficiently large. In this case, $\Delta_{t,c,\sigma}$ is considered to be infinite.}

\rev{Once $\Delta_{t,c,\sigma}$ and $\Delta_{t,c,jump}$ are known, we propose to define the model prediction for the critical tangential displacement, $\Delta_{t,c,mod}$, as the minimum value between $\Delta_{t,c,\sigma}$ and $\Delta_{t,c,jump}$ (see three possible different cases in Fig.~\ref{figmodel}). Doing so, any contact loaded by a $\Delta_t$ larger than $\Delta_{t,c,mod}$ will reach a fully sliding phase before contact separation (either through the transition to sliding, or after a jump instability in a compressive state). Thus, in the model, $\Delta_{t,c,mod}$ is indeed the seeked critical displacement separating the two adhesion regimes. Three possible cases are sketched in Fig.~\ref{figmodel}: when there is no intersection between $a_{jump}$ and $a_{\sigma}$ (case 3), $\Delta_{t,c,mod}=\Delta_{t,c,jump}$ ; when both $\Delta_{t,c,\sigma}$ and $\Delta_{t,c,jump}$ are finite, if  $\Delta_{t,c,jump} < \Delta_{t,c,\sigma}$ (case 1) then $\Delta_{t,c,mod}=\Delta_{t,c,jump}$ ; conversely, if  $\Delta_{t,c,jump} > \Delta_{t,c,\sigma}$ (case 2) then $\Delta_{t,c,mod}=\Delta_{t,c,\sigma}$.}

We now check to what extent the model can reach quantitative agreement with the measurements. For that, we need to estimate, from the experiments, the values of all parameters \rev{required to evaluate} $\Delta_{t,c,mod}$. $R$ is fixed by the geometry of the mold used to prepare the PDMS spheres. $E^*$ and $w_0$ are estimated from an unloading experiment with no preliminary shear (Tab.\ref{tab1}). \rev{The friction strength, $\sigma$, is estimated from the data in full sliding, as described in section~\ref{sec:prop} (see $\sigma_{unload}$ in Tab.\ref{tab1}). Estimating the value of $\alpha$ is less straightforward. In the literature, two distinct methods to access $\alpha$ exist, that have been reproduced here. $\alpha_{Pap}$ is estimated from the shear-induced area reduction, $A(Q)$, during the shearing phase (see section~\ref{sec:prop}). For all types of interfaces, we find values of $\alpha_{Pap}$ of order 0.01 (see Tab.~\ref{tab1}), compatible with those of~\cite{ciavarella_degree_2020} for PDMS. Alternatively, $\alpha_{Peng}$ is estimated from the value of the pull-off force of fully sliding contacts (see section~\ref{sec:prop}). The latter actually corresponds to the pull-off force deep in our second adhesion regime (on the right of the dashed vertical lines in Fig.~\ref{fig2}B, where $\Delta_{t}$ is larger than $\Delta_{t,s}$ and much larger than $\Delta_{t,c}$). For all types of interfaces, we find values of $\alpha_{Peng}$ of order 1 (see Tab.~\ref{tab1}), compatible with those of~\cite{peng_effect_2021} for PDMS.}

\rev{With all those values, we first estimate the value of $\Delta_{t,c,jump}$. In the absence of any argument in favor of one or the other evaluation of $\alpha$, we test both $\alpha_{Pap}$ or $\alpha_{Peng}$. While, with $\alpha_{Pap}$, $\Delta_{t,c,jump}$ is a few hundreds of micrometers for all types of interfaces (see line 1 in Tab.~\ref{tab2}), with $\alpha_{Peng}$, $\Delta_{t,c,jump}$ is only a few tens of micrometers (see line 2 in Tab.~\ref{tab2}). This discrepancy is fully attributable to the $\sim$100 ratio between $\alpha_{Pap}$ and $\alpha_{Peng}$. Explaining such a difference between both estimates of $\alpha$, already present in the related literature articles~\citep{ciavarella_degree_2020,peng_effect_2021}, is beyond the scope of the present experimental work. We emphasize however that, in the related models, $\alpha$ is merely a phenomenological way of accounting for the potentially complex way energy is dissipated at the molecular level of the contact, under mixed-mode loading. In the current state of knowledge, one cannot exclude that such dissipation might be different in a shearing experiment under constant normal load (the condition used to evaluate $\alpha_{Pap}$, see~\cite{ciavarella_degree_2020}) and in a pull-off test of a previously sliding contact (the condition used to evaluate $\alpha_{Peng}$, see~\cite{peng_effect_2021}), possibly explaining the difference between the values of $\alpha_{Pap}$ and $\alpha_{Peng}$.}

\begin{table}[hbt!]
\centering
\begin{tabular}{lccc}
Parameter & Standard/glass & Soft/glass & Soft/PMMA \\
\hline
\rev{$\Delta_{t,c,jump}$} predicted using $\alpha_{Pap}$ (\unit{\micro\meter}) & 133$\pm$24 & 282$\pm$26 & 233$\pm$22\\
\rev{$\Delta_{t,c,jump}$} predicted using $\alpha_{Peng}$ (\unit{\micro\meter}) & 10.7$\pm$0.9 & 28.7$\pm$2.1 & 24.7$\pm$2.5\\
\rev{$\Delta_{t,c,\sigma}$ predicted using $\alpha_{Pap}$ (\unit{\micro\meter})} & \rev{47$\pm$2} & \rev{162$\pm$10} & \rev{127$\pm$6}\\
\rev{$\Delta_{t,c,\sigma}$ predicted using $\alpha_{Peng}$ (\unit{\micro\meter})} & \rev{+$\infty$} & \rev{+$\infty$} & \rev{+$\infty$}\\
\hline
\rev{$\Delta_{t,c,mod}$ predicted using $\alpha_{Pap}$ (\unit{\micro\meter})} & \rev{47$\pm$2} & \rev{162$\pm$10} & \rev{127$\pm$6}\\
\rev{$\Delta_{t,c,mod}$ predicted using $\alpha_{Peng}$ (\unit{\micro\meter})} & \rev{10.7$\pm$0.9} & \rev{28.7$\pm$2.1} & \rev{24.7$\pm$2.5}\\
\hline
$\Delta_{t,c}$ observed from $A(z)$ (\unit{\micro\meter}) & $\in$[40 ; 60] & $\in$[40 ; 200] & $\in$[50 ; 200] \\
\end{tabular}
\caption{Comparison between observed and predicted $\Delta_{t,c}$. \rev{Lines 1 to 4: estimates of $\Delta_{t,c,jump}$ and $\Delta_{t,c,\sigma}$, using either $\alpha_{Pap}$ or $\alpha_{Peng}$. Lines 5 and 6: corresponding model predictions $\Delta_{t,c,mod}$. Line 7: observed $\Delta_{t,c}$. Error bars on model calculations} are standard deviations over 10000 evaluations of \rev{$\Delta_{t,c,jump}$ or $\Delta_{t,c,\sigma}$}, with $E$, $w_0$ , \rev{$\sigma$,} $\alpha_{Pap}$ and $\alpha_{Peng}$ drawn from gaussian distributions of mean (resp. standard deviation) the values (resp. error bars) from Tab.~\ref{tab1}.}\label{tab2}
\end{table}

\rev{The impact of the difference between $\alpha_{Pap}$ and $\alpha_{Peng}$ is also important, now even qualitatively, when considering the estimates of $\Delta_{t,c,\sigma}$. For $\alpha_{Pap}$, $\Delta_{t,c,\sigma}$ is of order several tens of micrometers, smaller than $\Delta_{t,c,jump}$ (case 2 in Fig.~\ref{figmodel}), for all types of interfaces (see line 3 in Tab.~\ref{tab2}). In striking contrast, for $\alpha_{Peng}$, $a_{\sigma}$ is found smaller than $a_{jump}$ whatever $\Delta_t$ (case 3 in Fig.~\ref{figmodel}), indicating that the contact would always undergo a jump instability, and not a transition to sliding. This is why infinite values are indicated in line 4 of Tab.~\ref{tab2}. Finally, for each case, the predicted critical tangential displacement, $\Delta_{t,c,mod}$, is taken as the minimum between $\Delta_{t,c,jump}$ and $\Delta_{t,c,\sigma}$. For all interfaces, $\Delta_{t,c,mod}=\Delta_{t,c,\sigma}$ when using $\alpha_{Pap}$ (line 5 in Tab.~\ref{tab2}), while $\Delta_{t,c,mod}=\Delta_{t,c,jump}$ when using $\alpha_{Peng}$ (line 6 in Tab.~\ref{tab2}).}

\rev{We now compare $\Delta_{t,c,mod}$ to the experimentally observed values of $\Delta_{t,c}$ (last line of Tab.~\ref{tab2}). When $\alpha_{Peng}$ is used, the predictions are systematically smaller than the observations, by a factor $\sim$2--5, a difference not accountable for by uncertainties. Instead, when $\alpha_{Pap}$ is used, the predictions agree with the observations, within the experimental error bars announced in Tab.~\ref{tab2}. Such an agreement is unexpected, given that most of the assumptions of the underlying model~\citep{papangelo_effect_2020}, including axisymmetry of the contact and the JKR limit,  are far from satisfied by our experimental system. In addition, while the quantitative agreement is striking in the case of standard-PDMS/glass interfaces, it is less satisfactory for interfaces based on soft-PDMS. Indeed, the large error bars provided for the observed $\Delta_{t,c}$ (up to \unit{200}{\micro\meter}, see line 7 in Tab.~\ref{tab2}), are mainly due to the obscuring effect of the oscillations in the corresponding $A(z)$ curves in Figs.~\ref{fig4}A and C. If one refers instead to Figs.~\ref{fig5}A and B, the actual value of $\Delta_{t,c}$ that separates the two adhesion regimes appears to remain in the range \unit{[40 ; 60]}{\micro\meter}, i.e. 2-3 times below the corresponding model predictions.}

\rev{Overall, the above first modelling attempt is promising because it provides realistic qualitative insights into the physical mechanisms potentially involved in the transition between the two observed adhesion regimes. Those mechanisms presumably involve a competition between the jump instability and the transition to sliding, controlled in particular by interfacial parameters like the reversible slip index, $\alpha$, the estimation of which remains challenging. As a matter of fact,  our quantitative tests of the model prediction of $\Delta_{t,c}$ remain essentially inconclusive: $\Delta_{t,c}$ is satisfactorily captured for only one of the three types of interfaces, and for only one of the two known ways of estimating $\alpha$.} We believe that identifying the origins of such \rev{a quantitative} failure would be an important step towards a full understanding of our empirical observations.

\rev{In this respect, an important observation is that finite strains are expected to play a role in the present experiments. As argued in~\cite{lengiewicz_finite_2020}, the characteristic interfacial shear strain in our contacts during sliding is $\sigma/G$, where $\sigma$ is the sliding friction stress of the interface and $G$ is the shear modulus of the PDMS (for such an incompressible material, $G=E/3$). Based on the values of Tab.~\ref{tab1}, $\sigma/G$ ranges from 0.57 for standard-PDMS/glass to 1.05 for soft-PDMS/glass interfaces. Those very large values indicate that finite strains are indeed expected in our experiments, especially those involving a fully sliding regime, further explaining why the linear-elasticity-based models tested above fail quantitatively. In this context, quantitative reproduction of our experimental results will likely involve finite strains, for instance within finite element frameworks like those of~\cite{mergel_continuum_2019,mergel_contact_2021,lengiewicz_finite_2020}.}

\subsection{Comparison with other types of contact interfaces and implications}
We eventually place our results in the context of previous studies of the effect of preliminary shear or sliding on pull-off forces, made on other tribological systems.

Recent nanoscale adhesion tests on hard materials yield contrasted results. For instance, \cite{milne_sliding_2019} report a significant sliding-induced increase of the pull-off force of silicon nanocontacts, an effect interpreted as the creation of covalent bonds after the sliding-induced removal of passivating surface species. Such a tribo-chemical phenomenon is not expected in our elastomer/glass experiments, possibly explaining the opposite behaviours. In contrast, \cite{sato_ultrahigh_2022} report a decrease of the pull-off force of silver nanocontacts with preliminary sliding, consistent with our results. Interestingly, the same above-discussed continuum models~\citep{papangelo_effect_2020,peng_effect_2021} also fail to quantitatively capture those nanotribology measurements, confirming the necessity to improve existing models of sheared adhesive contacts, irrespective of the length-scale.

Shear-induced control on adhesion is also key to the locomotion of climbing animals. Although adhesive pads vary among species (wet or dry, hairy or smooth), adhesion is often found an increasing function of the shear force when pulling the pad towards the body. Such adhesion enhancement presumably involves a combination of surface microstructures, shear-dependent peeling angle, slip-induced dissipation and specific rheological properties of contact-mediating secretions~\citep{labonte_biomechanics_2016,federle_dynamic_2019}. In this respect, our measurements on dry interfaces between smooth solids with well-defined macroscopic shape may help clarify the respective contributions of the above-mentioned mechanisms.

\section{Conclusion}
Overall, our results bring new experimental insights into the  fundamental and practical issue of the interplay between adhesion and friction, which resists the surface science community for decades~\citep{weber_experimental_2022}. Here, we identified a characteristic shear displacement, $\Delta_{t,c}$, that separates two different detachment regimes of dry soft and smooth contacts. From an empirical perspective, we showed that $\Delta_{t,c}$ is the minimum preliminary shear displacement necessary for the contact to reach full sliding before separation, irrespective of whether sliding occurs already during preliminary shearing or during the subsequent unloading. \rev{Better} predicting quantitatively the value of this critical shear displacement, and \rev{fully} unraveling the system parameters that control it, will \rev{likely} require the development of improved models. Practically, applying preliminary shears represents a new potential way to tune macroscopic adhesive forces in various applications, from haptics to soft robotics.

\section*{Acknowledgement}
We thank E. Delplanque for preparing PDMS samples, and M. Guibert, A. Aymard and A. Papangelo for discussions. This work was supported by LABEX iMUST (Grant ANR-10-LABX-0064) of Université de Lyon within the program “Investissements d’Avenir” (Grant ANR-11-IDEX-0007) operated by the French National Research Agency (ANR). It was also funded by ANR through Grant ANR-21-CE06-0048 (WEEL project). Since this research was funded, in whole or in part, by ANR, a CC-BY-NC-ND public copyright license has been applied by the authors to all previous versions of this document, up to the Author Accepted Manuscript arising from this submission, in accordance with the grant’s open access conditions.



\bibliographystyle{elsarticle-harv} 





\end{document}